\documentclass[12pt]{article}
\pdfoutput=1
\usepackage[latin1]{inputenc}

\usepackage{amsmath}
\usepackage{color}
\usepackage{amsfonts}
\usepackage{amssymb}
\usepackage{graphicx}
\usepackage{geometry}
\usepackage{amssymb,epsfig,subfigure}
\usepackage{hyperref}
\usepackage{comment}

\usepackage[numbers,sort&compress]{natbib}

\def\simge{
    \mathrel{\rlap{\raise 0.511ex 
        \hbox{$>$}}{\lower 0.511ex \hbox{$\sim$}}}}

\def\simle{
    \mathrel{\rlap{\raise 0.511ex 
        \hbox{$<$}}{\lower 0.511ex \hbox{$\sim$}}}}

\makeatletter
\renewcommand\section{\@startsection {section}{1}{\z@}%
                                 {-3.5ex \@plus -1ex \@minus -.2ex}
                                   {2.3ex \@plus.2ex}%
                                   {\normalfont\large\bfseries}}
\renewcommand\subsection{\@startsection{subsection}{2}{\z@}%
                                   {-3.25ex\@plus -1ex \@minus -.2ex}%
                                     {1.5ex \@plus .2ex}%
                                     {\normalfont\bfseries}}
\renewcommand\subsubsection{\@startsection{subsubsection}{3}{\z@}%
                                   {-3.25ex\@plus -1ex \@minus -.2ex}%
                                     {1.5ex \@plus .2ex}%
                                     {\normalfont\itshape}}
\makeatother

\def\pplogo{\vbox{\kern-\headheight\kern -29pt
\halign{##&##\hfil\cr&{\ppnumber}\cr\rule{0pt}{2.5ex}&\ppdate\cr}}}
\makeatletter
\def\ps@firstpage{\ps@empty \def\@oddhead{\hss\pplogo}%
  \let\@evenhead\@oddhead 
}
\def\maketitle{\par
 \begingroup
 \def\thefootnote{\fnsymbol{footnote}}
 \def\@makefnmark{\hbox{$^{\@thefnmark}$\hss}}
 \if@twocolumn
 \twocolumn[\@maketitle]
 \else \newpage
 \global\@topnum\z@ \@maketitle \fi\thispagestyle{firstpage}\@thanks
 \endgroup
 \setcounter{footnote}{0}
 \let\maketitle\relax
 \let\@maketitle\relax
 \gdef\@thanks{}\gdef\@author{}\gdef\@title{}\let\thanks\relax}
\makeatother

\numberwithin{equation}{section}

\newcommand{\nn}{\nonumber}

\renewcommand{\dag}{\dagger}

\newcommand{\M}{\mathcal{M}}

\newcommand{\be}{\begin{equation}}
\newcommand{\bea}{\begin{eqnarray}}
\newcommand{\ee}{\end{equation}}
\newcommand{\eea}{\end{eqnarray}}
\newcommand\beq{\begin{equation}}
\newcommand\eeq{\end{equation}}

\newcommand{\mc}{\mathcal}

\newcommand{\pp}{p_\parallel}

\newcommand{\qp}{q_\parallel}
\newcommand{\sv}{\text{sgn}(v)}
\newcommand{\Gr}{\Gamma_\text{reg}}
\newcommand{\Gs}{\Gamma_\text{sing}}

\def\be{\begin{equation}}
\def\ee{\end{equation}}
\def\ba#1\ea{\begin{align}#1\end{align}}
\def\bg#1\eg{\begin{gather}#1\end{gather}}
\def\bm#1\em{\begin{multline}#1\end{multline}}
\def\bmd#1\emd{\begin{multlined}#1\end{multlined}}

\def\nn{\nonumber}

\def\({\left(}
\def\){\right)}
\def\[{\left[}
\def\]{\right]}

\begin{document}

\setcounter{page}0
\def\ppnumber{\vbox{\baselineskip14pt
}}

\def\ppdate{
} \date{}

\author{A. Liam Fitzpatrick$^{1,2}$, Gonzalo Torroba$^3$ and Huajia Wang$^1$\\
[7mm] \\
{\normalsize \it $^{1}$Stanford Institute for Theoretical Physics, Stanford 
University, Stanford, CA 94305, USA} \\
{\normalsize \it $^2$SLAC National Accelerator Laboratory, 2575 Sand Hill Road, 
Menlo Park, CA 94025, USA }\\
{\normalsize \it $^3$Centro At\'omico Bariloche and CONICET, R8402AGP Bariloche, 
Argentina}
}

\bigskip
\title{\bf  Aspects of Renormalization in \\ Finite Density Field Theory
\vskip 0.5cm}
\maketitle

\begin{abstract}
We study the renormalization of the Fermi surface coupled to a massless boson near 
three spatial dimensions. For this, we set up a Wilsonian RG with independent decimation procedures for bosons and fermions, where the four-fermion interaction ``Landau parameters'' run already at tree level. Our explicit one loop analysis resolves previously found obstacles in the renormalization of finite density field theory, including logarithmic divergences in non-local interactions and the appearance of multilogarithms. The key aspects of the RG are the above tree level running, and a UV-IR mixing between virtual bosons and fermions at the quantum level, which is responsible for the renormalization of the Fermi velocity.
We apply this approach to the renormalization of $2 k_F$ singularities, and a companion paper considers the RG for Fermi surface instabilities. We end with some comments on the renormalization of finite density field theory 
with the inclusion of Landau damping of the boson.
\end{abstract}
\bigskip

\newpage

\tableofcontents

\vskip 1cm

\section{Introduction}\label{sec:intro}

Quantum field theory (QFT) at finite density appears as the continuum limit in a 
wide array of systems, ranging from neutron stars to novel quantum critical 
points in condensed matter physics. Its quantum properties are, however, much 
less well understood than in the relativistic case.  In particular, additional 
subtleties arise when the Fermi surface is coupled to gapless modes. In the 
present work we focus on the renormalization of such theories, with the goal of 
setting up a consistent renormalization group (RG) approach.

Some of the most interesting applications of finite density occur at strong 
coupling, e.g. as models of strongly-correlated electron systems; however, it 
has been hard to identify a general framework where finite density field 
theories at strong coupling can be understood. On the other hand --and we will 
see this explicitly below-- weakly coupled limits already exhibit very 
interesting physics, and challenge standard RG ideas. From experience in 
relativistic theories, understanding in detail the perturbative limits is a key 
to the strong coupling problem. Indeed, one can try to identify properties of 
finite density field theory at weak coupling that extend to the more general 
case, for instance by use of an $\epsilon$ expansion or in certain large $N$ 
limits.

In this work we will study the renormalization of finite density field theory at 
weak coupling and in a small $\epsilon$ expansion, continuing the analysis 
started in~\cite{Mahajan:2013jza,Fitzpatrickone,Torroba:2014gqa}. The class of 
theories that we will focus on contains massless bosons (scalars or gauge 
fields) coupled to fermions at finite density by a Yukawa interaction of the 
form $g \phi \psi^\dag \psi$. We will work near $d=3$ spatial dimensions, the 
upper critical dimensional where the coupling $g$ is marginal and a weak 
coupling expansion can be set up.

The coupled system of a critical boson interacting with a Fermi surface is not 
fully understood even at weak coupling, and requires going beyond the Fermi 
liquid RG set up in~\cite{shankar, Polchinski:1992ed} . The main new effect of 
finite density is that bosons and fermions have very different energetics and 
scalings. Low energy bosonic excitations occur near $\vec k=0$; therefore, boson 
momenta scale towards the origin at low energies. In contrast, IR fermionic 
excitations occur at the Fermi surface of finite momenta $\vec k=\vec k_F$. When these 
two sets of degrees of freedom are allowed to interact, novel quantum effects 
arise, which are absent in the relativistic theory or in models with only 
fermions at finite density. 

Our task is to determine a consistent RG for
the coupled system of bosons and finite density fermions. Important first steps were taken in developing such an RG in the seminal works on color superconductivity in QCD at finite density \cite{Son:1998uk,Shuster}.\footnote{This approach has recently been applied to theories of bosons coupled to the Fermi surface in \cite{Metlitski}.}  A key result in that approach, which we make use of as well, is the existence of logarithmic divergences in 4-Fermi interactions already at 
tree level. However, going beyond the leading order analysis in \cite{Son:1998uk, Shuster}, many new subtleties arise.  The main goal of this paper is to develop a fully systemic Wilsonian approach that may be used to higher orders and that addresses the non-trivial issues that arise there.  In particular, our proposal will resolve two problems that made the renormalization of systems of coupled bosons and finite density fermions quite 
involved: the nonlocal (singular) logarithmic divergences recently found 
in~\cite{Torroba:2014gqa}, and the presence of multilogs in 4-Fermi interactions from 
exchange of massless bosons.  We will show the tree-level running of four-fermi interactions explicitly in the calculation of the Wilsonian 
action,  
together with the 
required tree level counterterms. In terms of these running couplings we will 
argue that the theory is renormalizable.  Applications of this RG to instabilities of the Fermi surface will be presented in the companion paper~\cite{ustoappear}.

First, in \S \ref{sec:treelevel} we present the RG approach, explaining the decimation procedures, the origin of UV-IR mixing in the presence of massless bosons, and the tree level running of the Landau parameters. Next, \S \ref{sec:quantum} contains our main results at one loop. We prove the cancellation of divergences in all (marginally) relevant interactions, and compute the RG beta functions. These results are then applied in \S \ref{sec:appl} to the study of the RG flow for the Landau parameters and for the $2k_F$ vertex. Lastly, \S \ref{sec:higher} briefly discusses the possible extension to higher loop order, with the inclusion of Landau damping of the boson. We end by discussing interesting future directions in \S \ref{sec:future}. In the Appendix we perform a detailed analysis of the theory in dimensional regularization, a regulator that is efficient in correctly capturing the various quantum effects and that will also be needed in future higher loop extensions of our work.

\section{Scaling and renormalization at tree level}\label{sec:treelevel}

In this first section we will discuss some general properties of renormalization 
of quantum field theories where bosons interact with a finite density of 
fermions. 
Applying the renormalization group to finite density theories is quite 
challenging for several reasons. First, the scaling properties of bosons and 
fermions are now very different: low energy bosonic excitations are located near 
the origin of momentum space, while for fermions they occur around the Fermi 
surface $\vec k = \vec k_F$.  Secondly, the Fermi surface leads to enhanced 
quantum contributions from the large number of light degrees of freedom. This 
enhancement can compensate for the phase suppression factor from momentum 
integration, with the result that low energy excitations will also contribute to 
the RG evolution of the theory. The Wilsonian RG of Shankar and Polchinski~\cite{shankar, Polchinski:1992ed}, where fermion shells of 
high momentum are integrated out, needs to be modified. Our task is to set 
up an RG that can deal consistently with these problems.

\subsection{Classical theory}\label{subsec:classical}

For concreteness we will consider a real spin zero boson interacting with a 
spinless fermion,
\be\label{eq:Sphi1}
S= \int d\tau\, d^dx \,\left\{\frac{1}{2} \left((\partial_\tau \phi)^2+c^2 (\vec 
\nabla \phi)^2 \right)+ \psi^\dag \left(\partial_\tau+ \varepsilon_F(i\vec \nabla) -\mu_F
\right)\psi +L_{int}\right\}
\ee
where $\varepsilon_F(\vec k)$ is the quasiparticle energy, and $\mu_F$ is the chemical potential.
The basic property of the fermion energy function $\varepsilon(\vec k)\equiv \varepsilon_F(\vec k)-\mu_F$ is 
that it admits a Fermi surface at finite momentum, $\varepsilon(\vec k_F)=0$.  For instance, $\varepsilon(\vec k) = \vec k^2/2m -\mu_F$ for a massive fermion at finite chemical potential $\mu_F$.
Our analysis can be easily extended to other fields, such as gauge bosons, Dirac 
fermions, etc. Given this matter content, possible interaction terms include a 
boson-fermion Yukawa coupling, as well as 4-Fermi and $\phi^4$ couplings,
\be
L_{int} =g \phi \psi^\dag \psi+ \lambda (\psi^\dag \psi)^2+  \lambda' \phi^4 + 
\ldots
\ee
These interactions can depend on momenta, and we will define them in more detail shortly.
The classical $\phi^4$ coupling will not play an important role, so we set $ 
\lambda'=0$. 

In order to study the low energy theory near the Fermi surface in its simplest 
form, it will be convenient to
assume a spherical Fermi surface. Following the spherical RG of~\cite{shankar, 
Polchinski:1992ed}, the fermion momentum is measured radially from the Fermi 
surface,
\be\label{eq:kdecomp}
\vec k = \hat n (k_F + k_\perp)\,,
\ee
with $\hat n$ a unit vector perpendicular to the Fermi surface. In the low 
energy theory $\omega \ll \mu_F$, $k_\perp \ll k_F$, it is sufficient to expand
\be\label{eq:Sf}
S_f= \int d\tau \frac{d \Omega_n}{(2\pi)^{d-1}} 
\frac{dk_\perp}{2\pi}\,\psi^\dag(\vec k) (\partial_\tau+ v k_\perp+ 
\frac{w}{2k_F} k_\perp^2 + \ldots) \psi(\vec k)\,.
\ee
Here we have rescaled the fermion to absorb an overall power of $k_F$.

We will see that in the effective theory at energies much smaller than the Fermi energy $\mu_F$, 
quadratic and higher order corrections to the dispersion relation can be 
neglected. It will then be sufficient to keep the leading linear term, working 
with a flat band of Fermi velocity $v= \varepsilon_F'(\vec k_F)$. In this case, 
our results will also apply locally on the Fermi surface, even if the surface is 
not spherical (as long as it is smooth). This local approximation, which will be useful below, should not be confused with the patch RG (see e.g.~\cite{Polchinski1994, Nayak1994a}), where normal and tangential directions in the patch have different scalings.

\subsection{Renormalization approach}\label{subsec:prescription}

\begin{figure}[h!]
\begin{center}
\includegraphics[width=0.55\textwidth]{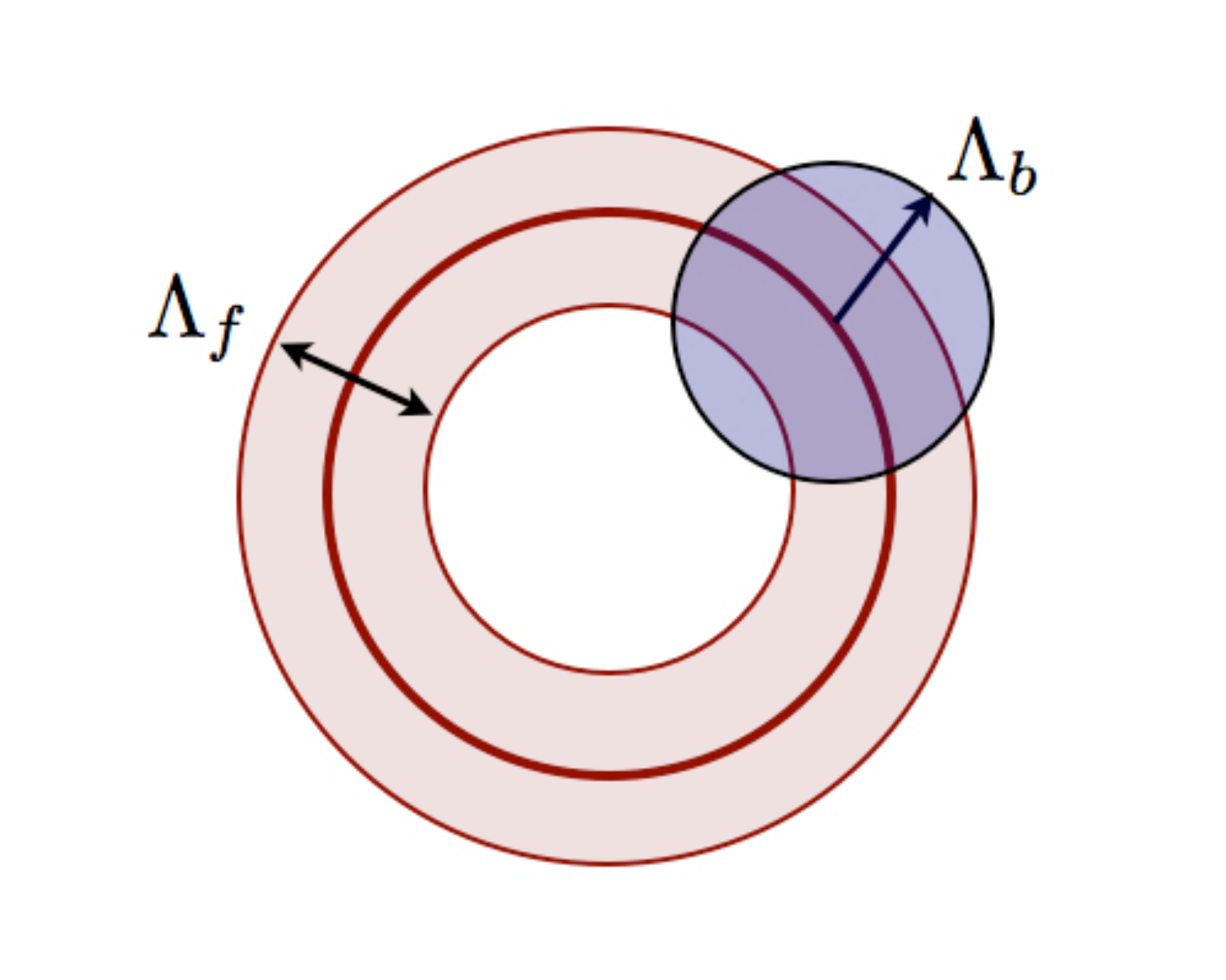}
\end{center}
\caption{\small{Scaling towards low energies for bosons (left) and fermions 
(right). Low energy bosonic modes occur near the origin, while for fermions they 
arise around the Fermi surface. }}\label{fig:two-cutoffs}
\end{figure}

Because of the very different boson and fermion scalings, it is natural to 
define an effective theory that depends on two independent cutoffs $\Lambda_b$ 
and $\Lambda_f$. One cutoff $\Lambda_b$ regulates the scaling of the bosonic 
degrees of freedom towards the origin, $p^2 < \Lambda_b$; the other cutoff 
$|\varepsilon(p)|<\Lambda_f$ dictates the displacement from the Fermi surface 
in the low energy theory.  The two scales can be varied independently, and as a 
result it is possible to have different decimation procedures for bosons and 
fermions. This is illustrated in Figure \ref{fig:two-cutoffs} for a spherical 
Fermi surface.

A smooth version of these cutoffs is obtained by modifying the propagators of 
bosons and fermions in terms of a function $K(x)$,
\be
D(p)= \frac{K(\vec p^{\,2}/\Lambda_b^2)}{p_0^2+ \vec p^{\,2}}\;,\;G(p)= - 
\frac{K(\varepsilon(\vec p)^2/\Lambda_f^2)}{i p_0 - \varepsilon(\vec p\,)}
\ee
with $K(x) \to 1$ for $x \ll 1$ and $K(x) \to 0$ for $x \gg 1$.
We furthermore choose regulators that do not constrain the frequency, which will be integrated over the whole range $(-\infty, \infty)$.\footnote{ One could alternately regulate the frequency range as well.  Some Feynman integrals in the theory are finite and yet still dependent on the ratio of the frequency cut-off to momentum cut-off, and so different choices correspond to different choices for local counter-terms.  We caution the reader that this can lead to additional scheme-dependence of $\beta$ functions at higher loops beyond what one might perhaps be used to.}
Under an infinitesimal variation of the cutoffs $\delta 
\Lambda_{b,f}$, the action changes in order to keep the partition function 
fixed~\cite{Wilson:1973jj, Polchinski:1983gv}, defining a bidimensional RG flow.
Having these two cutoffs will also allow us to distinguish between local and 
nonlocal renormalization effects, as well as the origin of multilogs~\cite{ustoappear}. A tree level RG for the coupled boson-fermion system similar to that suggested in Fig. \ref{fig:two-cutoffs} was studied in~\cite{yamamoto}. It is 
interesting to note that two-dimensional RG flows also appear for jets in QCD, 
see e.g.~\cite{QCD, Chiu:2012ir}.

In the effective theory with the two cutoffs $\Lambda_b$ and $\Lambda_f$, fermions can scatter with a maximum exchanged momentum set by $\Lambda_b$. As a result, it will be sufficient to restrict to a local patch (or two antipodal patches) of the Fermi surface of angular size $\Lambda_b/k_F$, as shown in Figure \ref{fig:two-cutoffs}. We will return to this point below in \ref{subsec:treelevel}. A basic approximation that we will use throughout this work is that we will linearize the fermion dispersion relation, effectively neglecting quadratic and higher order terms in $\varepsilon(\vec p)$. This will be valid over the whole patch as long as
\be\label{eq:cutoff-constraint}
\Lambda_b \lesssim \sqrt{2 k_F \Lambda_f}\,,
\ee
a relation that we will assume in what follows.\footnote{We are assuming that $v$ and $w$ are comparable in (\ref{eq:Sf}), as is the case for an approximately quadratic dispersion relation. Otherwise an extra factor of $v/w$ is required in (\ref{eq:cutoff-constraint}).}

A problem with momentum cutoffs is that they can break Ward identities and hence 
may require additional counterterms. In our case a finite density analog of the 
Ward identity of $U(1)$ charge	 will play an useful guiding role, and we wish to 
preserve it. For this reason, in the renormalization of the theory it can be convenient to use a dimensional regularization (DR) procedure that preserves gauge invariance, taking the limit $\epsilon \to 0$ in $d=3-\epsilon$.  
However, due to its 
conceptual transparency, we will mostly use the hard cut-offs in the body of the paper, and describe renormalization using DR in appendix \ref{app:dimreg}.

Computationally, the theory will be renormalized following the standard QFT 
procedure of writing the bare fields and couplings (denoted by a subindex `$0$') 
in terms of physical quantities plus counterterms. For the action 
(\ref{eq:Sphi1}), the required redefinitions are
\be\label{eq:renorm1}
\psi_0 = Z_\psi^{1/2} \psi\;,\;\phi_0= Z_\phi^{1/2}\phi\;,\;v_0= Z_v 
v\;,\;g_0=\mu^{\epsilon/2}\,\frac{Z_g}{Z_\phi^{1/2}Z_\psi}g\;,\;\lambda_0 = 
\frac{Z_\lambda}{Z_\psi^2}\lambda\,,
\ee
where $\mu$ is an arbitrary RG scale and $g$ and $\lambda$ are dimensionless. 
The counterterms $Z_i$ are adjusted to cancel divergences, and the beta 
functions are calculated noting that the bare parameters are $\mu$ independent. 
Below we will work in terms of counterterms $\delta_i$ defined as
\be\label{eq:deltaidef}
Z_\psi=1+\delta_\psi\;,\;Z_v\,v=v+\delta v\;,\;Z_g\,g=g+\delta g\;,\;Z_\lambda 
\, \lambda=\lambda+\delta \lambda\,.
\ee

\subsection{UV-IR mixing}

Let us explain, at this general level, how the Fermi surface leads to 
enhanced quantum contributions from low energy excitations. These effects will 
play an important role in the renormalization of the velocity and the 
boson-fermion cubic coupling. For this, we note that close to the Fermi surface, 
the product of two fermion propagators with momenta $p$ and $p+q$ has the 
structure
\be\label{eq:GG}
G(p) G(p+q) \approx G(p)^2 +2\pi \,\frac{\sv q_\perp}{iq_0- v q_\perp}\,\delta(p_0) 
\delta(p_\perp)\,,
\ee
in the limit $q\rightarrow 0$ and in the cutoff regulator that we just introduced.\footnote{We note that the integral $\int_p G(p) G(p+q)$ is ambiguous in the low energy theory and it depends on the ratio of the frequency and momentum cutoffs. Here we are taking the frequency cutoff to infinity first.} 
This can be seen by integrating both sides over $p_0$ and $p_\perp$. The 
consequences of this for the scattering of quasiparticles and properties of 
collective excitations are well known; see e.g. the discussion in~\cite{AGD}. 

In Fermi liquids without gapless bosons, the singular second term in (\ref{eq:GG}) does 
not contribute to the RG and is not included. 
However, when the system is coupled to a gapless boson, such singular terms in 
the product of fermion lines can lead to logarithmic enhancements from the 
exchange of virtual boson. An important example is 
the one loop correction to the cubic vertex, proportional to
\be
\int_p \,D(p) G(p) G(p+q)\,.
\ee
The second term in (\ref{eq:GG}), when multiplying the boson propagator, 
produces an extra logarithmic divergence,
\be\label{eq:additionaldiv}
\int_p\,D(p)\,\frac{q_\perp}{iq_0- v q_\perp}\,\delta(p_0) \delta(p_\perp) 
\sim\,\frac{q_\perp}{iq_0- v q_\perp}\,\int\, \frac{d^2 
p_\parallel}{p_\parallel^2}\,,
\ee
which has to be taken into account in the renormalization of the 
theory.\footnote{It is interesting to note that, similarly to 
(\ref{eq:additionaldiv}), in the SCET theory of QCD there are also additional 
logs from collinear divergences.} This is a UV-IR mixing, where low energy fermionic excitations can exchange high energy bosonic modes, contributing, as we shall see, to the running of couplings in the effective theory.

\begin{figure}[h!]
\begin{center}
\includegraphics[width=0.70\textwidth]{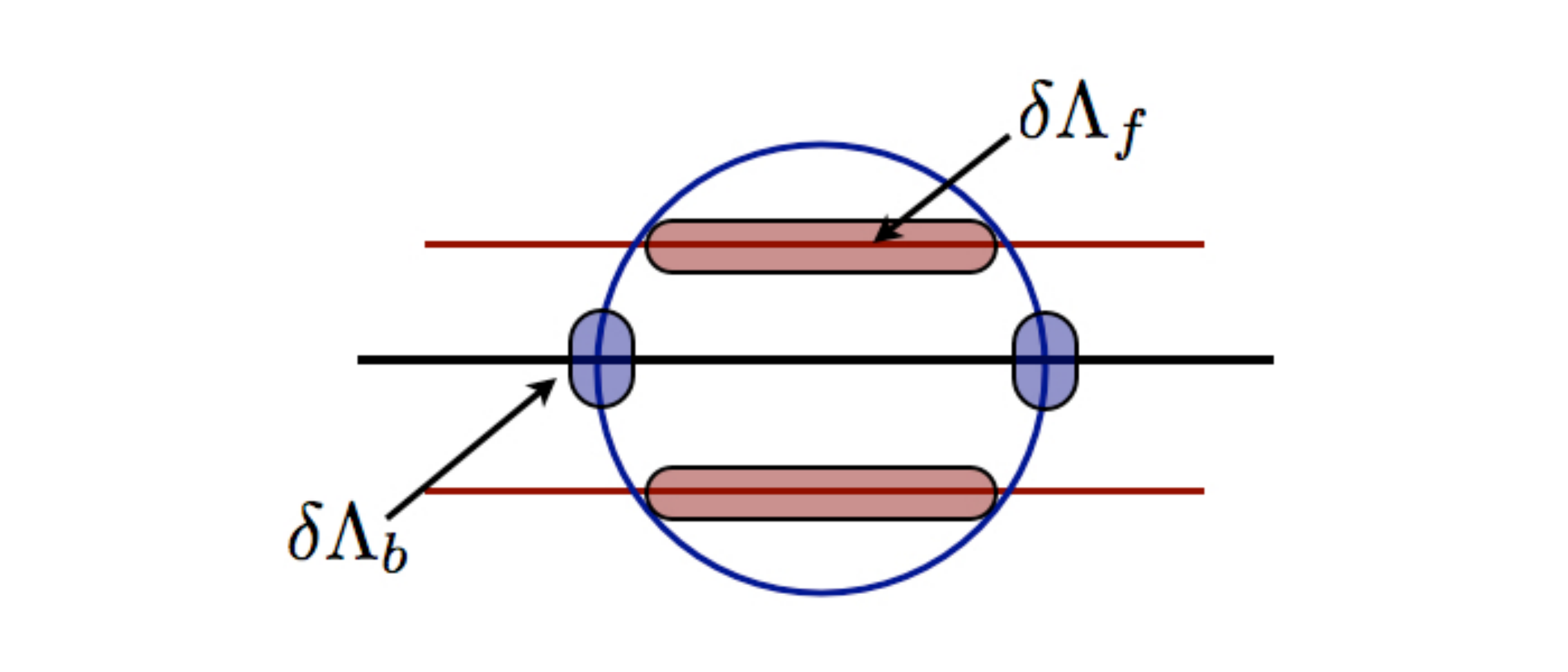}
\end{center}
\caption{\small{Dominant contributions to the RG for the Fermi surface coupled 
to a gapless boson. A local patch of the Fermi surface (here shown in black) is 
delimited by $\Lambda_b$, while the width of low energy excitations around the 
surface is controlled by $\Lambda_f$. By varying these cutoffs, the dominant 
quantum contributions come from the two shells in the Figure determined by 
$\delta \Lambda_f$ and $\delta \Lambda_b$.}}\label{fig:RGstep}
\end{figure}

As a result, we will find that quantum effects are dominated by two regions of 
momenta near the Fermi surface, as shown in Figure \ref{fig:RGstep}. One class 
of logs will come from the standard RG towards the Fermi surface and will be 
regulated by $\Lambda_f$. The other contributions will be proportional to $\log 
\Lambda_b$, and will be generated by virtual bosons as in 
(\ref{eq:additionaldiv}); these effects appear due to the delta function peak in 
the product of two fermion lines at the Fermi surface.

\subsection{Tree level RG for 4-Fermi interactions}\label{subsec:treelevel}

The first contribution to the RG occurs 
already at tree level, in the form of a logarithmic running of 4-Fermi 
interactions due to the exchange of virtual bosons \cite{Son:1998uk,Shuster,Metlitski}. We will now discuss in 
detail how this comes about, which will be the basis for renormalization at the loop level.

Let us first recall some properties of 4-Fermi interactions, which we write as
\be\label{eq:4Fermigeneral}
S \supset -\int \prod_i d^{d+1}k_i\, 
\lambda(k_4,\ldots, k_1)\,\delta^{d+1}(k_1+k_2-k_3-k_4)\,\psi(k_1) \psi(k_2) \psi^\dag(k_3) 
\psi^\dag(k_4)\,.
\ee
Our convention is that $\lambda<0$ represents an attractive interaction.
Only two kinematic configurations are marginal in the presence of a Fermi 
surface: forward scattering (`FS'), where the angle between the incoming pair of fermions is 
the same as that for the outgoing pair; and BCS, where the initial angles are 
opposite~\cite{shankar}. 
In $d=2$, the forward scattering constraint has the simple form $\theta_1= 
\theta_3$ and $\theta_2= \theta_4$, or its permutation. Then the coupling 
becomes a function $\lambda( \theta_1 - \theta_2)$ of the relative angle between 
the incoming fermions. Instead, $\theta_1=-\theta_2$ and $\theta_3=-\theta_4$ 
for BCS, with the coupling now being a function $\lambda(\theta_1-\theta_3)$.

In the spherical RG, the angle on the Fermi surface plays the role of a flavor index for the low energy excitations. It is then convenient to write this coupling function in a 
basis of spherical harmonics,
\be\label{eq:Un}
\lambda_L= \int_0^{2\pi}\,\frac{d\theta}{\sqrt{2\pi}}\,e^{i L \theta}\, 
\lambda(\theta)\,,
\ee
which will have simple properties under 
renormalization. (Here $\theta=\theta_{12}$ or $\theta_{13}$ for forward scattering and BCS, respectively). The generalization to $d=3$ uses Legendre polynomials, and 
$\cos \theta_{ij}= \vec k_i \cdot \vec k_j/k_F^2$. In a local analysis on the 
Fermi surface, one can equivalently choose a basis of plane waves.  Such a local 
analysis will typically be appropriate for our  renormalization of the forward 
scattering four-fermion interactions.  We will see in a moment that this is because the exchange of the 
massless boson at momenta below the cut-off $\Lambda_b$ corresponds to 
scattering between nearby (i.e. nearly collinear) directions on the Fermi 
surface, with a change in angles of order $\Lambda_b/k_F$.  For this reason, and 
also for simplicity, in what follows, we will use such a plane wave basis.

Consider now the effect of
integrating out a shell of high momentum bosons; this generates an effective 
$\psi^4$ interaction
\be\label{eq:4Fermi1}
S\supset - \frac{g^2}{2} \int_{q,p,p'}\, \frac{\delta K(q^2/\Lambda_b^2)}{q_0^2 + 
\vec q^{\,2}} \psi^\dag(p+q) \psi(p) \psi^\dag(p'-q) \psi(p')\,,
\ee
where the variation $\delta K$ of the smooth cutoff function has the effect of 
restricting $q$ to a shell of size $\delta \Lambda_b$. In the notation of (\ref{eq:4Fermigeneral}), this generates a 4-Fermi coupling
\be
\lambda(p'-q,p+q,p',p)=-\frac{g^2}{2}\frac{\delta K(q^2/\Lambda_b^2)}{q_0^2 + 
\vec q^{\,2}} ,
\ee
and the FS and BCS marginal channels correspond to
\be
\text{FS}:\,\,p \approx p'-q\;,\;\text{BCS}:\,\,p \approx -p'\,.
\ee
Notice that the fermion momenta $p$ and $p'$ are of order $k_F$, while the exchanged boson momentum $|\vec q| \sim \Lambda_b \ll k_F$. This effective 4-Fermi interaction then couples patches of the Fermi surface over small angular sizes $\sim \Lambda_b/k_F$ (nearly tangential patches for FS, and antipodal for BCS). As we anticipated above, this is the reason why we can restrict to a local analysis on the Fermi surface over angular scales $\sim \Lambda_b/k_F$.

The question we need to 
address is how to absorb this momentum-dependent vertex into a renormalization 
of constant couplings. This is accomplished by changing to the angular momentum basis, where the modes change by
\be
\delta \lambda_L = g^2 \int \,\frac{d^2 q_\parallel}{(\sqrt{2\pi})^2}\frac{e^{i \vec{L} \cdot 
\vec{q}_\parallel/k_F }}{q_0^2 + q_\perp^2+q_\parallel^2} \delta 
K(q^2/\Lambda_b^2)  \approx  g^2\, J_0(|L|\Lambda_b/k_F) \frac{\delta 
\Lambda_b}{\Lambda_b} .
\ee
The Bessel function $J_0$ decays and oscillates for large $|L|$ and thus 
logarithmic running will effectively be turned off exponentially quickly for 
$\Lambda_b |L| \gg k_F$.

\begin{figure}[h!]
\begin{center}
\includegraphics[width=1.0\textwidth]{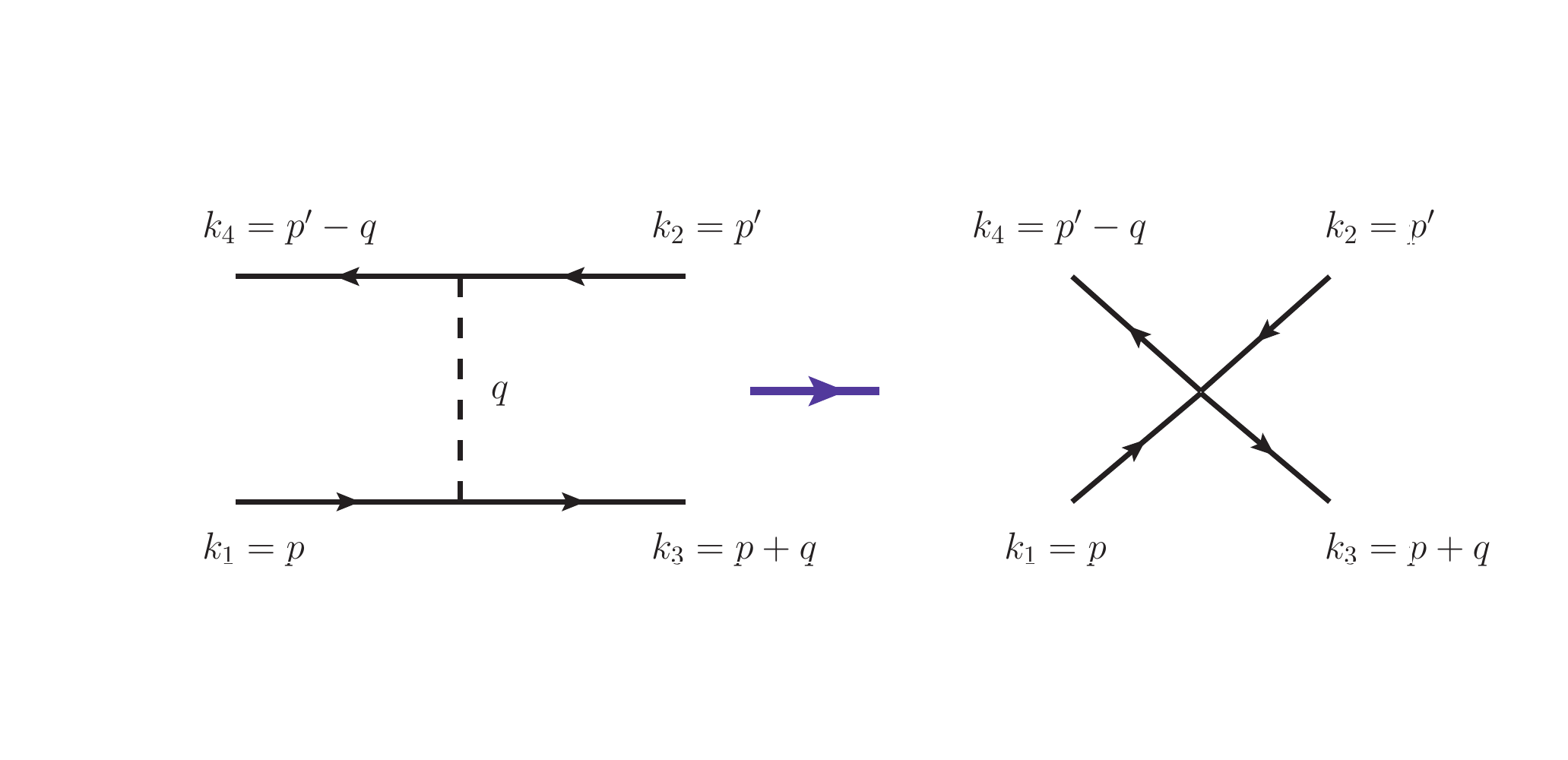}
\end{center}
\caption{\small{Boson-mediated $\psi^4$ interaction at tree 
level. The FS channel is $p \approx p'-q$, and the BCS channel is $p \approx -p'$.}\label{fig:psi4}}
\end{figure}

This is depicted in Figure \ref{fig:psi4}. It is important to stress that in the 
effective 4-Fermi interaction (\ref{eq:4Fermi1}), the exchanged momentum $q$ is 
restricted to a shell of high momenta. This in turn constrains the angular 
integration in the spherical harmonic decomposition to $\Delta \theta= 
\Lambda_b/k_F<1$, which is the origin of the $\log \Lambda_b$ dependence that we 
just encountered.

Therefore, already at tree level we need to add a counterterm for $(\psi^\dag 
\psi)^2$ in order to cancel this divergence and have a finite physical coupling. 
We write the bare coupling in terms of the 
physical coupling plus a counterterm:
\be
\lambda_{0,L} = \lambda_L+ \delta \lambda_L ,
\ee
and choose
\be\label{eq:deltalambda}
\delta \lambda_L = g^2 \,\log \Lambda_b
\ee
for $|L| < k_F/\Lambda_b$. Higher modes do not run.
Equivalently, one can work in dimensional regularization (DR). Since this is a 
mass-independent scheme, there is no upper bound on the $|L|$ for which 
$\lambda_L$ gets a log divergence and thus $\delta \lambda_L = 
g^2/\epsilon$ for all $\vec{L}$.  It requires taking $k_F\rightarrow \infty$ from the beginning, before taking $\epsilon\rightarrow 0$, and thus formally $|L|/k_F$ is always below the RG scale $\mu$;  this is fine as long as one restricts to $|L| < k_F/\mu$. 
We conclude that there is a tree level running of the $\psi^4$ coupling,
\be\label{eq:beta0}
\beta_{\lambda_L} = -\frac{d\lambda_L}{d \log \Lambda_b}=g^2\,.
\ee
An attractive coupling 
($\lambda<0$ in our convention) then grows towards the IR.  Our analysis has been for general fermion momenta $p$ and $p'$, so the tree level beta function (\ref{eq:beta0}) applies to both FS and BCS channels of 4-Fermi interactions.

The existence of a tree level beta function for $\lambda_L$ is special to finite 
density. This does not occur in zero density relativistic theories, although 
effects such as that of Figure \ref{fig:psi4} are formally generated by a  
Wilsonian exact RG~\cite{Polchinski:1983gv}. Even in the exact Wilsonian 
RG, in the relativistic case one can always eliminate such tree-level running 
terms by an extra step where their effects are reproduced by loop-level terms. In our context, the tree level logarithmic enhancement is a consequence of 
exchanging high momentum bosons that induce virtual scatterings tangential to 
the Fermi surface, so that modes far below the cut-off can exchange a 
mode near the cut-off. Since the diagram in the figure is not one-particle 
irreducible, this is an example of a situation where the 1PI action and the 
Wilsonian action are different.
In the following sections such tree level 
contributions will be shown to be key for understanding the RG of the quantum 
theory.

\section{Renormalization of the quantum theory at one loop}\label{sec:quantum}

In this section we prove the renormalizability of the theory at one loop, and 
calculate the RG beta functions. This requires evaluating the one loop 
corrections to the boson and fermion self-energies, as well as the 
renormalization of the $\phi \psi^\dag \psi$ and $(\psi^\dag \psi)^2$ vertices. 
The vacuum polarization of the boson is finite and has been calculated in detail 
in many places (see e.g.~\cite{Hertz1976, Millis1993,Nagaosa}), and will not be repeated here. We will return to its effects in higher loop diagrams in \S \ref{sec:higher}.

We will find that the one loop RG is scheme-dependent, both on the regularization (cutoff versus DR) and on the renormalization subtraction. Scheme dependence in finite density QFT starts at one loop due to the presence of tree level running --in the relativistic context, this would usually be seen only from two loops (and in theories with a single coupling, only from three loops).
For computational purposes, it is always more convenient to work in DR, and this is indeed the scheme where we originally found many of the effects that will be described shortly. However, it turns out that the standard minimal subtraction scheme in DR leads to unphysical results for the RG (such as wrong sign anomalous dimensions), because it misses important running effects in the quadratic action from finite pieces. For this reason, in this section we will present the results mostly with the cutoff regulator of \S \ref{subsec:prescription} and in a physical subtraction scheme. Dimensional regularization of the finite density theory will be studied in the Appendix. We will also check that both regulators agree on a physical subtraction, as it should be.

\subsection{Ward identities near the Fermi surface}

The field theory has a conserved charge associated to complex rotations of the 
fermions, which becomes a gauge symmetry if a dynamic electromagnetic field is added. The expression for the current in the low energy theory is
\be\label{eq:Jmu}
J_\mu(p) = \psi^\dag(p) V_\mu \psi(p)+\mathcal O(p_\perp/k_F)\,.
\ee
We have introduced the velocity 4-vector 
\be
V_\mu \equiv (i, -v \hat n)
\ee
such that $V\cdot p= i p_0 - v \hat n \cdot \vec p$ is the fermion kinetic term 
in the effective theory. Varying the direction $\hat n = \vec p/|\vec p|$ 
obtains an infinite number of approximately conserved conserved currents, each 
associated to the effective theory on a small Fermi surface patch (this neglects 
curvature corrections and interpatch couplings).

Quantum-mechanically, the conserved current leads to a Ward identity relating 
the 3-point function $\langle J_\mu \psi^\dag \psi\rangle$ to the fermion 
self-energy --see e.g.~\cite{Weinberg:1995mt}. In more detail, defining
\be
\langle J_\mu(q) \psi^\dag(p+q) \psi(p)\rangle= \Gamma_\mu(p;q)\,G(p)\,G(p+q)\,,
\ee
the Ward identity reads
\be\label{eq:Ward1}
q_\mu \Gamma_\mu(p;q)= G^{-1}(p)-G^{-1}(p+q)\,.
\ee
Here $G$ is the full fermion propagator, related to the tree level expression 
and the self-energy by 
\be
G^{-1}=G_0^{-1}-\Sigma\,.
\ee 
In a relativistic theory, 
taking the limit $q \to 0$ leads to the infinitesimal version of the Ward 
identity
\be
\Gamma_\mu= \partial_\mu \Sigma\,.
\ee
However, we will see that at finite density the limits $q_0 \to 0$ and $| \vec 
q\,| \to 0$ do not commute; this makes the infinitesimal identity ambiguous, but 
(\ref{eq:Ward1}) is still well-defined. Similar ambiguities are found in QCD at finite density~\cite{Brown:2000eh,Schafer:2004zf}.

Eq.~(\ref{eq:Ward1}) can be used to relate the quantum corrections to the Yukawa 
vertex\footnote{Our convention is that $(g+ \Gamma)\phi \psi^\dag \psi$ gives 
the renormalized vertex.}
\be
\Gamma(p;q) = - \langle \phi(q) \psi^\dag(p+q) \psi(p) \rangle-g
\ee
and the fermion self-energy. In the approximation (\ref{eq:Jmu}), we have $\Gamma_\mu \approx 
V_\mu \Gamma/g$, and plugging this into (\ref{eq:Ward1}) then obtains
\be\label{eq:Ward2}
\Gamma(p;q)= g \frac{\Sigma(p+q) - \Sigma(p)}{V \cdot q}+\mathcal 
O(q_\perp/k_F)\,,
\ee
a result that will be important for our renormalization approach.

\subsection{Vertex renormalization and fermion self-energy}\label{subsec:vertex}

\begin{figure}[h!]
\begin{center}
\includegraphics[width=0.70\textwidth]{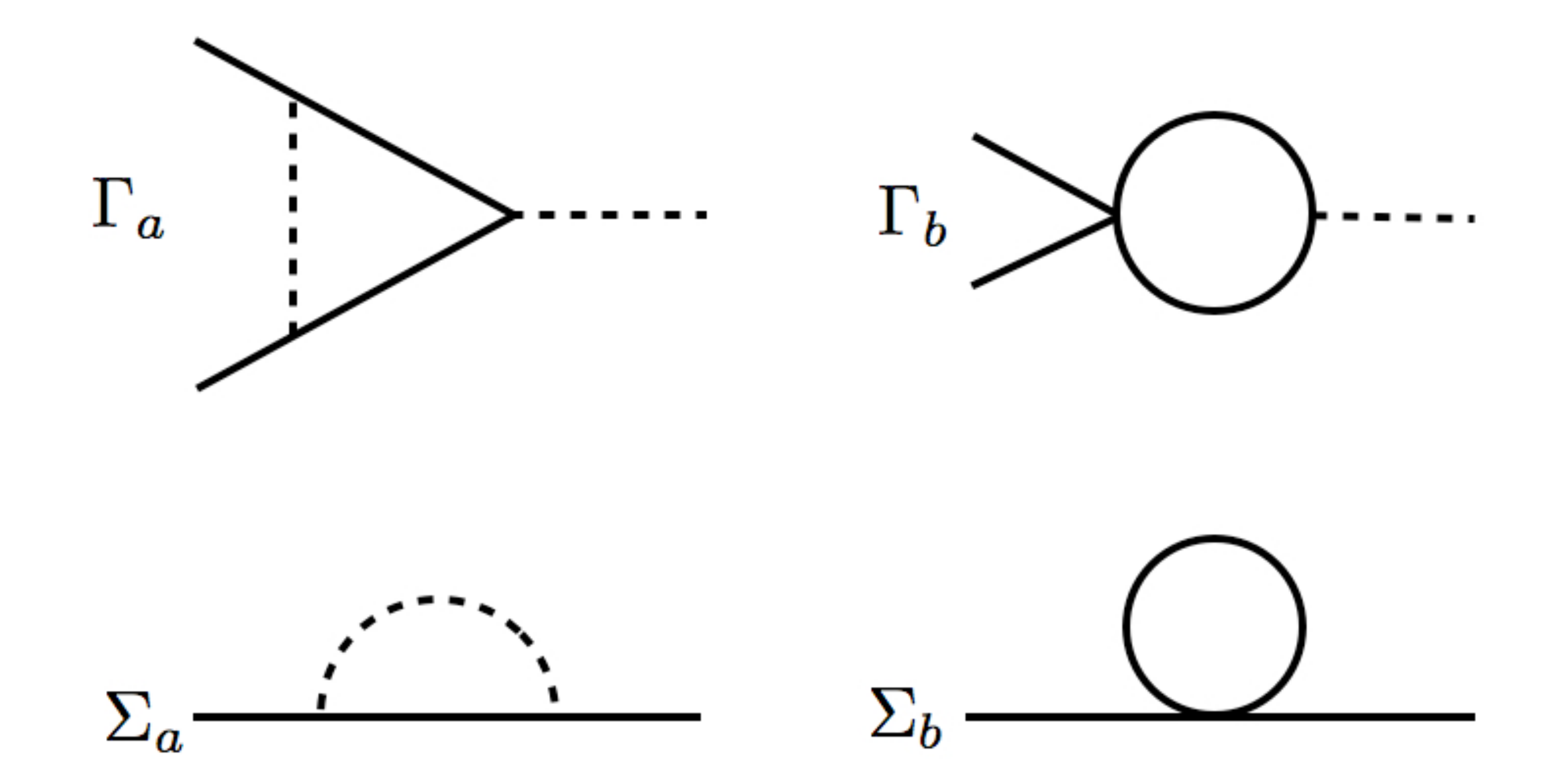}
\end{center}
\caption{\small{One-loop diagrams that renormalize the cubic vertex and fermion 
self-energy.}\label{fig:diagrams1}}
\end{figure}

Let us now calculate the one loop corrections to the fermion self-energy and 
Yukawa vertex, shown in Figure \ref{fig:diagrams1}. This analysis will clarify the 
origin of singular operators generated at the quantum level. In our calculations 
we use renormalized perturbation theory [see (\ref{eq:renorm1})], and adjust the counterterms in order to cancel loop divergences.

\subsubsection{Vertex correction}

The one loop vertex correction $\Gamma_a$ of Figure \ref{fig:diagrams1} is, 
linearizing the fermion propagators around the Fermi surface,
\be\label{eq:Vertex_correction}
\Gamma_a(k;q 
)\approx\frac{g^3}{(2\pi)^{4}}\int\frac{dp_0\,dp_\perp\,d^{2}
p_\parallel}{(k_0-p_0)^2+(k_\perp - p_\perp)^2+p_\parallel^2}\,\frac{1}{ip_0-v 
p_\perp}\,\frac{1}{i(p_0+q_0)-v(p_\perp+q_\perp)}\,.
\ee
The momenta have been decomposed with respect to the external fermion momentum 
$\vec k = \hat n(k_F+ k_\perp)$ as follows:
$\vec p = \hat n(k_F+ p_\perp)+ \pp$, and $\vec q= \hat n q_\perp + \qp$. 

There are two types of contributions to the vertex correction. First there is a 
\textit{regular} piece $\Gr$ that comes from evaluating (\ref{eq:Vertex_correction}) 
setting $q=0$ inside the integral. This is dominated by the region of large momenta around $\Lambda_f$
in Fig. \ref{fig:RGstep}. However, there 
is an additional divergence, that is not accounted for in the usual 
treatment~\cite{Torroba:2014gqa}. This \textit{singular} piece $\Gs=\Gamma-\Gr$ comes from a small region 
$p_\perp+ q_\perp \approx 0$ around the origin where the two fermion poles are 
on opposite sides of the real axis. This is the contribution from 
the delta function peak on the Fermi surface from the product of 
two fermion lines, that we discussed before in (\ref{eq:GG}). Normally such terms would not contribute to the RG, but in our model they are logarithmically enhanced due to the exchange of virtual bosons, and should be included.\footnote{Another model where singular contributions play a role in the renormalization of the theory is \cite{chubukov1}.}

The regular and singular contributions to the vertex depend on the regularization scheme, and here we evaluate these using the cutoff prescription set up in \S\ref{subsec:prescription}. The DR expressions are given in the Appendix. In
\be
\Gr=\frac{g^3}{(2\pi)^{4}}\int\frac{dp_0\,dp_\perp\,d^2
p_\parallel}{p_0^2+p_\perp^2+p_\parallel^2}\,\frac{1}{(ip_0-v 
p_\perp)^2}
\ee
we have to integrate over $p_0$ first; integrating then over $p_\parallel$ (this is convergent and can be extended to $\Lambda_b \to \infty$) and finally over $p_\perp$ with cutoff $\Lambda_f$ yields
\be\label{eq:Gr1}
\Gr=\Gamma(\hat n k_F;0)\approx \frac{g^3}{4\pi^2}\frac{\log\Lambda_f}{1+|v|}\,,
\ee
up to finite terms.
The singular contribution comes 
the delta function peak on the Fermi surface from the product of 
two fermion lines, that we discussed before in (\ref{eq:GG}). It evaluates to
\be\label{eq:Gs1}
\Gs(\hat n k_F;q) \approx \frac{g^3}{4\pi^2}\,\frac{\sv q_\perp}{iq_0- 
v q_\perp}\,\log \Lambda_b\,.
\ee

Eq.~(\ref{eq:Gs1}) is quite puzzling 
because it contains a logarithmic divergence for a singular\footnote{By singular (sometimes also referred to as `nonlocal') 
we mean that the operator has a singular dependence on frequency/momenta.} 
operator. At face value, this would imply that the theory is not renormalizable. 
One could argue that the nonlocal term is not Wilsonian --it comes from the 
delta function peak in (\ref{eq:GG}) right on the Fermi surface-- and hence 
should not be included in the RG. However, ignoring logs is dangerous: in the IR 
such effects grow and the Wilsonian prediction will not be a good approximation 
to the physical answer. Moreover, in higher loop diagrams nonlocal divergences 
will mix with local ones and, again, these effects cannot be ignored. 
Physically, we have a UV-IR mixing, where an IR contribution from excitations at 
the Fermi surface multiplies a UV enhancement from high momentum modes.

However, we will now show that the solution lies in the second diagram 
$\Gamma_b$ of Figure \ref{fig:diagrams1}, which will precisely cancel the 
nonlocal divergence. This diagram has not been included in previous RG 
approaches, since the fermion loop from the forward scattering 4-Fermi 
interaction is finite --in fact, it vanishes if the fermion is restricted to a 
high momentum shell as in~\cite{shankar}. In the present case, however, the 
situation is different because already at tree level the 4-Fermi interaction has 
a logarithmic divergence. Inserting this into the one loop diagram, and taking 
into account the delta function peak in (\ref{eq:GG}), will give the right 
contribution to make the renormalization procedure well-defined. The reason behind this cancellation is that the region of fermion momenta that gives rise to the nonlocal renormalization is the same one where boson exchange leads to the effective 4-Fermi vertex through the process of Figure \ref{fig:psi4}.

Let us see how this comes about.
Denoting the momenta in the external fermion lines by $p$ and $k$, so that there 
is an incoming boson momentum $k-p$, diagram $\Gamma_b$ in Figure 
\ref{fig:diagrams1} gives
\be
\Gamma_b(p,k;k-p)=-g \int d^2 L (\lambda_L+ \delta \lambda_L)  
\,\int\,\frac{d^{d+1}q}{(2\pi)^{d+1}}\, e^{i \vec{q}_\parallel \cdot 
\vec{L}/k_F}G(k+q) G(p+q)\,,
\ee
where the overall minus sign comes from the fermion loop. The 4-Fermi vertex is in the FS channel, but we will avoid writing explicitly the `FS' subscript in this section, as this is the only type of vertex that will appear.\footnote{In the applications in \S \ref{sec:appl} the BCS vertex will also enter, and then we will distinguish this channel explicitly.}

The $d^2 
\vec{q}_\parallel$ integration is done trivially since the fermion propagators 
depend only on $q_\perp$, and only the zero mode $\lambda_{L=0}$ 
contributes. The remaining integration is finite but ambiguous, which also reflects the scheme dependence that we found before. In the regularization with $\Lambda_b$ and $\Lambda_f$, the integral over frequencies is done first as in (\ref{eq:GG}), obtaining
\be
\Gamma_b(p, k; k-p)=- (\lambda_{L=0} + \delta \lambda_{L=0})g\,\frac{\sv}{4\pi^2}\,\frac{k_\perp-p_\perp}{i(k_0-p_0)- v (k_\perp-p_\perp)}\,.
\ee
Combining this with $\Gamma_a$ in (\ref{eq:dimreg_gamma}), and using 
$\delta \lambda_{L=0}=g^2\,\log \Lambda_b$ for the counterterm [see 
(\ref{eq:deltalambda})], the nonlocal divergences precisely cancel and we arrive 
to
\be\label{eq:Gammatotal}
\Gamma(p;q)\approx\frac{g^3}{4\pi^2}\frac{1}{1+|v|}\,\log \Lambda_f-\frac{
g\lambda_{L=0}}{4\pi^2}\,\frac{\sv\,q_\perp}{iq_0- v q_\perp}\,.
\ee
Note that the divergence in the singular contribution 
(\ref{eq:Gs1}) is a $\log \Lambda_b$. This illustrates the more general point 
that the seemingly nonlocal renormalization effects found in \cite{Torroba:2014gqa} are controlled by $\Lambda_b$.
The running velocity below 
will be another example of this.

\subsubsection{Fermion self-energy}

The one loop fermion self-energy turns out to be the most delicate renormalized quantity. First, with a cutoff regulator the result for $\Sigma_a$ is very sensitive to the way in which the fermion loop momentum is regulated. This problem is fortunately avoided in DR, and we present the corresponding calculation in the Appendix. Additionally, by power-counting there is a linear divergence in $\Sigma_b$ 
that complicates the regularization of the subleading logarithmically divergent 
kinetic terms. This affects both the cutoff and dimensional regulators.

The diagram $\Sigma_a$ is evaluated in detail in the Appendix, where we also show explicitly that it satisfies the Ward identity that relates it to $\Gamma_a$. We will then not repeat the analog calculation with cutoffs, and simply deduce $\Sigma_a$ from (\ref{eq:Gr1}) plus (\ref{eq:Gs1}) using (\ref{eq:Ward2}):
\be
\Sigma_a(q)=\frac{g^3}{4\pi^2}\frac{\log\Lambda_f}{1+|v|}(iq_0-vq_\perp)+\frac{g^3}{4\pi^2}\,\log \Lambda_b\,\sv q_\perp\,.
\ee

Next, the diagram $\Sigma_b(k)$ in Figure \ref{fig:diagrams1} is proportional to 
$\lambda_{L=0} \int dq_0 dq_\perp G(k+q)$. This integral is ambiguous in the low 
energy theory where the linear dispersion relation is used, and we need to 
regularize it in a way consistent with the Ward identity. 
As alluded to above, the computation of $\Sigma_b(k)$ is complicated by the linear 
divergence.  As this may be completely absorbed by a counter-term, we can 
without loss of generality compute the less-divergent quantity 
$\Sigma_b(k)-\Sigma_b(0)$:
\be
\Sigma_b(k) - \Sigma_b(0) \propto \lambda_{L=0} \int dq_0 \int 
d q_\perp \left( \frac{1}{i (q_0 + k_0) - v(q_\perp + 
k_\perp)} - \frac{1}{i q_0  - v q_\perp } \right).
\label{eq:dSigma}
\ee

In the cutoff approach we are instructed to perform the convergent $q_0$ integration first and obtain 
\be
\int d q_\perp \left( \Theta(q_\perp + k_\perp) - \Theta(q_\perp) \right) = 
k_\perp.
\ee
However, it is easy to see by inspection that the $dq_\perp$ integrand in  
(\ref{eq:dSigma}) vanishes if we shift the $q_\perp$ integration variable of 
$\Sigma_b(k)$ by $-k_\perp$ relative to $\Sigma_b(0)$.  In fact, 
(\ref{eq:dSigma}) has an ambiguity that is exactly parameterized by this 
relative shift $a_\perp$ in the integration variable $q_\perp$:
\begin{eqnarray}
\Sigma_b(k) - \Sigma_b(0) &\propto& \lambda_{L=0} \int dq_0 \int 
d q_\perp \left( \frac{1}{i (q_0 + k_0+a_0(k)) - v(q_\perp + 
k_\perp+a_\perp(k))} - \frac{1}{i q_0  - v q_\perp } \right).\nonumber\\
&\propto& \lambda_{L=0} \int d q_\perp \left( \Theta(q_\perp + k_\perp+a_\perp(k)) - 
\Theta(q_\perp) \right)  =\lambda_{L=0}( a_\perp(k) + k_\perp)\,.
\label{eq:dSigma2}
\end{eqnarray}
An identical ambiguity arises in the evaluation of triangle diagrams for 
anomalies in relativistic theories.  Here, as there, the parameter $a_\perp(k)$ 
represents an additional piece of data that must be input into the theory, 
either from matching to a UV theory or by constraints on the low-energy theory.  Taylor expanding $a_\perp(k)$ in $k$, we can discard the constant piece since $\Sigma_b(0)-\Sigma_b(0)=0$, and furthermore on dimensional grounds we should discard terms of ${\cal O}(k_0^2, k_\perp^2)$ and higher as well.  We are left with $a_\perp(k) = a_1 k_0 + a_2 k_\perp$ and hence
\be\label{eq:ambiguity1}
\Sigma_b(k)=-\frac{\lambda_{L=0}+\delta \lambda_{L=0}}{4\pi^2}\,\sv\,((a_1 k_0 + a_2k_\perp) +  k_\perp))\,,
\ee
For a given regulator, there is a unique choice for $a_1,a_2$ that respects gauge invariance.
In the present case, we see that the Ward identity (\ref{eq:Ward2}) holds only for $a_\perp(k)=0$. Thus,
\begin{eqnarray}\label{eq:Sigmab}
\Sigma_b(k)&=&-\frac{\lambda_{L=0}+\delta \lambda_{L=0}}{4\pi^2}\,\sv\,k_\perp\,,\\
\label{eq:Sigmatotal}
\Sigma(k)&\approx&\frac{g^2}{4\pi^2}\frac{\log \Lambda_f}{1+|v|}\,(ik_0- 
vk_\perp  )-\frac{\lambda_{L=0}}{4\pi^2}\,\sv\, k_\perp\,.
\end{eqnarray}

The first term in $\Sigma(k)$ is a wavefunction renormalization; notice, however, that the 
logarithmic divergences for velocity renormalization have cancelled between $\Sigma_a$ 
and $\Sigma_b$, with the result that the velocity renormalization is finite and 
proportional to $g \lambda_{L=0}$. This is analogous to the cancellation of the 
nonlocal terms in the vertex. Therefore, 
unlike~\cite{Fitzpatrick:2013rfa,Torroba:2014gqa}, we find no UV divergence for 
the velocity after taking into account the tree level running of the 4-Fermi vertex. 

This however highlights an important point about the running of terms in the quadratic action, specifically the velocity and the wavefunction renormalization.  Below, we will discuss how a physical subtraction scheme for the fermion self-energy {\it does} produce a running velocity.  More generally, it guarantees that the full physical logarithmic enhancement of $q_0$ and $q_\perp$ terms in $\Sigma$  gets taken into account in the running velocity $v$ and wavefunction factor $Z$, and therefore get fully included in the fermion propagator.  This is a significant advantage of physical subtraction;  the alternative would require calculations of physical amplitudes to include a large number of diagrams with insertions of the $\Sigma_b$ diagram from Figure \ref{fig:diagrams1} as subdiagrams, which would quickly become very difficult.  For the wavefunction factor, this is even more important, because its running produces anomalous dimensions that feed into the running of all parameters.  Thus, in practice one should always define the wavefunction renormalization counterterm $\delta Z$ so that it agrees with physical subtraction, and  we believe it is vastly easier and more transparent to define the running of $v$ this way as well.

\subsection{Comments on singular operators}\label{subsec:nonlocal}

Having calculated the one loop contributions to the fermion self-energy and 
cubic vertex, it will be useful now to clarify more the origin of the singular 
contributions that we have found.

Focusing on the self-energy (\ref{eq:Sigmatotal}), the first term corresponds to 
wavefunction renormalization $Z_\psi$, while the second term amounts to a 
correction to the velocity. We can combine both into a single momentum-dependent 
renormalization factor $\mathcal Z(k)$, rewriting the quantum effect as
\be
L \supset \mathcal -\mathcal Z(k) \psi^\dag(i k_0 - v k_\perp) \psi
\ee
with
\be\label{eq:mZ}
\mathcal Z(k) 
=\frac{g^2}{4\pi^2}\frac{\log \Lambda_f}{1+|v|}-\frac{\lambda_{L=0}}{4\pi^2}\, 
\frac{\sv\,k_\perp}{i k_0 - v k_\perp}\,.
\ee
Therefore, from the point of view of the original fermion dispersion relation,
the velocity renormalization corresponds to a singular contribution to $\mathcal Z(k)$.
We stress that the running of $v$
is a direct consequence of the UV-IR mixing discussed above: the IR 
enhancement of excitations near the Fermi surface multiplying a UV contribution 
from the exchange of virtual bosons tangential to the Fermi surface. Without these effects, the RG would be analytic and the velocity wouldn't run.

By the Ward identity, the factor $\mathcal Z(k)$ is the same as the vertex 
correction $\Gamma$, the correlation function that originally displayed the 
nonlocal contributions. The key point is that these effects are controlled by 
the physical 4-Fermi coupling $\lambda_L$ and do not require introducing a new 
coupling or counterterm, something that would have obstructed the renormalization of 
the theory. Moreover,
in terms of the two RG scales introduced in \S \ref{subsec:prescription}, their 
running is set by $\log \mu_b$, the boson scale.

\begin{figure}[h!]
\begin{center}
\includegraphics[width=0.70\textwidth]{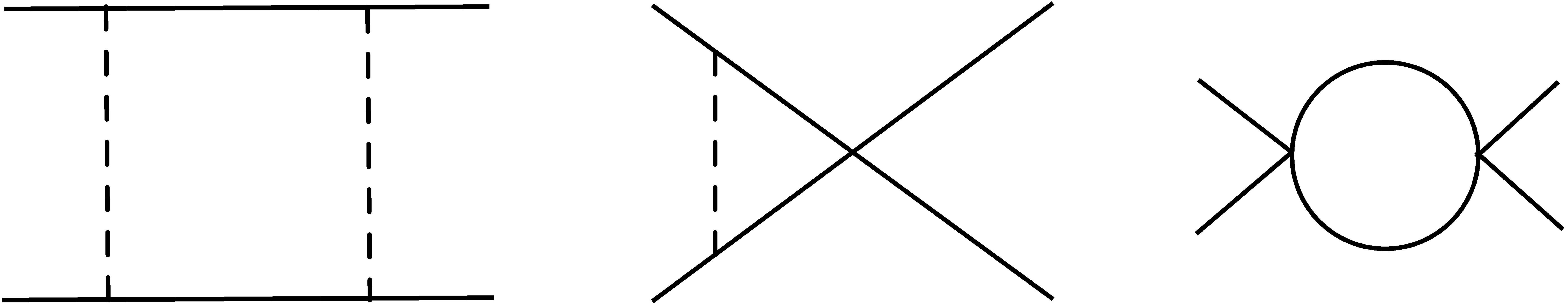}
\end{center}
\caption{\small{One loop contributions to the zero-sound (forward scattering) 
vertex.}}\label{fig:zero-sound}
\end{figure}

The poles in (\ref{eq:mZ}) or (\ref{eq:Gammatotal}) contain information about 
the physical spectrum of excitations near the Fermi surface. Similar 
singularities arise in the 4-Fermi vertex in the Fermi liquid, and should be 
treated in the same way~\cite{AGD}. The difference here is the logarithmic 
enhancement from boson exchange, reflected in the running of the physical 
coupling $\lambda_L$. Let us illustrate this with the set of one loop 
contributions to fermion scattering shown in Figure 
\ref{fig:zero-sound}.

The diagram from boson exchange gives a $\log^2 \Lambda_b$, which is cancelled 
by the insertion of the counterterm $\delta \lambda_L$ in the other two 
diagrams, with the result that the sum of the three diagrams depends only on the 
physical coupling $\lambda_L$.
Denoting the two fermion momenta for the forward scattering channel by $p$ and 
$k$ obtains
\be
\Gamma^{(4)}(p,k)=-\lambda^2 \,\int\,\frac{d^{d+1}q}{(2\pi)^{d+1}}\, G(k+q) 
G(p+q)= - \frac{\lambda^2}{4\pi^2}\,\frac{\sv\,(p_\perp-k_\perp)}{i(p_0-k_0)- v 
(p_\perp-k_\perp)}\,. 
\ee
This gives a singular contribution to the zero sound (forward scattering) 
vertex, which is familiar from quantum treatments of Fermi liquids~\cite{AGD}. 

Our point again is that,  this singular contribution is still controlled by the local coupling $\lambda(\mu_b)$ 
and does not induce an independent RG flow. Had we not included the tree level 
running of $\lambda$, the cancellation of logarithmic divergences would have 
failed, and we would have found a $\log^2$ multiplying a singular 
momentum-dependent function. The physical content here is the same as that for 
the renormalization of the velocity and the singular contribution to the cubic 
vertex, all of them being related by the Ward identity. It is not consistent to 
ignore these singular contributions, by the same reason that we cannot ignore 
the renormalization of the velocity.

\subsection{RG beta functions}

We are now ready to put these results together and determine the RG flow of the 
theory at one loop.\footnote{It will be useful to recall our sign conventions 
for the quantum corrections and counterterms. The corrections to the fermion 
kinetic term appear in the combination $\delta_\psi(ip_0-vp_\perp)- \delta v 
\,p_\perp + \Sigma$, while for the cubic vertex we have $g+\delta g+ \Gamma$.} We stress again that in finite density QFT one generically expects scheme dependence already at one loop, due to the tree level running of the 4-Fermi coupling. We have seen this before in the differences between the dimensional and cutoff regulators, and below additional scheme dependence will arise from the renormalization conditions.

Let us adopt a physical renormalization scheme where the 
renormalized couplings are defined in terms of physical amplitudes at an RG 
scale $\mu$. For simplicity we also scale the two cutoffs in the same way and denote them by $\Lambda$; we briefly discuss the bidimensional RG below. Defining $t=\log \mu/\Lambda$, from  (\ref{eq:Sigmatotal}) we read off the counterterms
\be
\delta_\psi= \frac{g^2}{4\pi^2(1+|v|)}\,t\;,\;\delta v= 
-\frac{\sv\,\lambda_{L=0}}{4\pi^2}
\ee
and recalling that $\beta_{\lambda_L}=g^2$ obtains the anomalous dimension and running velocity
\be\label{eq:beta1}
\gamma_\psi=\frac{g^2}{8\pi^2(1+|v|)}\;,\;\beta_v=\frac{\sv g^2}{4\pi^2}\,.
\ee
On the other hand, for the vertex correction (\ref{eq:Gammatotal}) we may 
define $g$ as the amplitude $\Gamma(p,q)$ evaluated at the point 
\be
q_0 = x \mu\,,
q_\perp=\mu\,.
\ee 
Then, we identify
\be
\delta 
g=\frac{g^3}{4\pi^2(1+|v|)}\,t-\frac{g\lambda_{L=0}}{4\pi^2}\,\frac{\sv}{v}\
\,\frac{1}{1-i x/v}\,.
\ee
In calculating $\beta_g$, the first term in $\delta g$ cancels against the 
anomalous dimension; however, the second term in $\delta g$ now gives a nonzero 
contribution proportional to $\beta_{\lambda_{L=0}}$. The $\beta$ function for the theory in $d=3-\epsilon$ becomes
\be\label{eq:betag}
\beta_g(x)= - \frac{\epsilon}{2} 
g+\frac{g}{4\pi^2}\,\frac{\sv}{v}\,\frac{1}{1-ix/v} \beta_{\lambda_{L=0}}= - 
\frac{\epsilon}{2} g+\frac{g^3}{4\pi^2|v|}\,\frac{1}{1-ix/v}\,.
\ee
We are free to choose any value for the ratio $x$ at the physical subtraction point; for instance the choice $x=\infty$ removes the last term.  

The $x$-dependent beta function (\ref{eq:betag}) may look puzzling at first, 
but we should stress that this is fairly generic of physical subtraction schemes. In fact, it can arise even in relativistic $\lambda \phi^4$ theory, where choosing a physical subtraction scheme of say $\M_{2 \rightarrow 2}(s_0,t_0,u_0)= \lambda$ for the $2\rightarrow 2$ amplitude $\M_{2\rightarrow 2}$ at some subtraction point $s_0,t_0,u_0$ would lead to a dependence of the $\beta$ function for $\lambda $ on the ratio of Mandelstam variables $s_0/t_0, s_0/u_0$ at sufficiently high loop order. This just reflects that in a physical subtraction scheme, one by definition subtracts off from the bare couplings $\lambda_B = \lambda + \delta \lambda$ any difference between the amplitude $\M(s_0, t_0, u_0)$ and the renormalized coupling $\lambda$ itself.  Since the finite part of the amplitude depends on $s,t$ times the running couplings, in order to do such a subtraction one is forced to subtract off a non-trivial running function of the subtraction point.  In other words, this encodes the fact that the finite part of the amplitude contains non-trivial dependence on dimensionless kinematic ratios (here, $s/t$ or $q_0/q$) times running couplings.  

It is generally stated in the literature that for the abelian theory in $d=3$, the beta function $\beta_g$ vanishes identically due to the Ward identity. However, in this case this statement should be interpreted with some care. Given the renormalization of the velocity, the Ward identity implies a beta function (\ref{eq:betag}) that depends on the ratio $x=q_0/q_\perp$ at the subtraction point. In the regime where $x \to \infty$, the beta function does vanish, but more generally $\beta_g(x) \neq 0$. So one must be precise about what limits of momenta one uses to define the running coupling if one wishes to obtain certain features of the $\beta$ function.  

We may also interpret these results using the bidimensional RG of \S 
\ref{subsec:prescription}. In this case, the nonsingular RG evolution is 
controlled by $\mu_f$, while singular effects evolve along $\mu_b$:
\be
\frac{\partial v}{\partial t_b}=\frac{\sv g^2}{4\pi^2}\;,\;\frac{\partial 
g}{\partial t_b}=\frac{g^3}{4\pi^2}\,\frac{\sv}{ix - v}\,.
\ee
We expect that at higher loop order the RG flows along $t_b$ and $t_f$ will 
start to mix. Furthermore, in generalizations of our theory with matrix-valued $\phi$, the 
anomalous dimension and $\delta g$ contributions will not cancel, with the 
result that $g$ evolves along both RG directions simultaneously. Of course, we 
can always project this bidimensional RG flow down to $t_b \propto t_f$, but 
having the two different scales helps to track the physical origin of the 
different quantum contributions.

\section{Applications}\label{sec:appl}

So far we have analyzed the renormalization of the Fermi surface coupled to a gapless boson, focusing on the anomalous dimension, running velocity, and vertex corrections. These operators can be defined and studied in a local patch, and are approximately insensitive to global properties of the Fermi surface.\footnote{Recall that the effective theory keeps bosons of momenta $|p| < \Lambda_b$, so the angular size of one of these patches is set by $\Lambda_b/k_F$.} There are, however, other observables of interest, where the quasiparticles can exchange large momenta of order $k_F$. In this section we will briefly consider the renormalization of these quantities, applying the approach described in \S\S \ref{sec:treelevel} and \ref{sec:quantum}.

One of the most important operators of this type is the 4-Fermi BCS interaction. In the presence of the massless scalar, this leads to a parametric enhancement in the condensation of Cooper pairs~\cite{Son:1998uk,Shuster}. The RG approach to Fermi surface instabilities will be discussed in detail in~\cite{ustoappear}. Here we will instead consider the renormalization of Landau parameters in \S \ref{subsec:Landau} and the $2 k_F$ vertex in \S \ref{subsec:2KF}, which are closely related to our analysis in the previous sections.

\subsection{Renormalization of Landau parameters}\label{subsec:Landau}

\begin{figure}[h!]
\begin{center}
\includegraphics[width=1.0\textwidth]{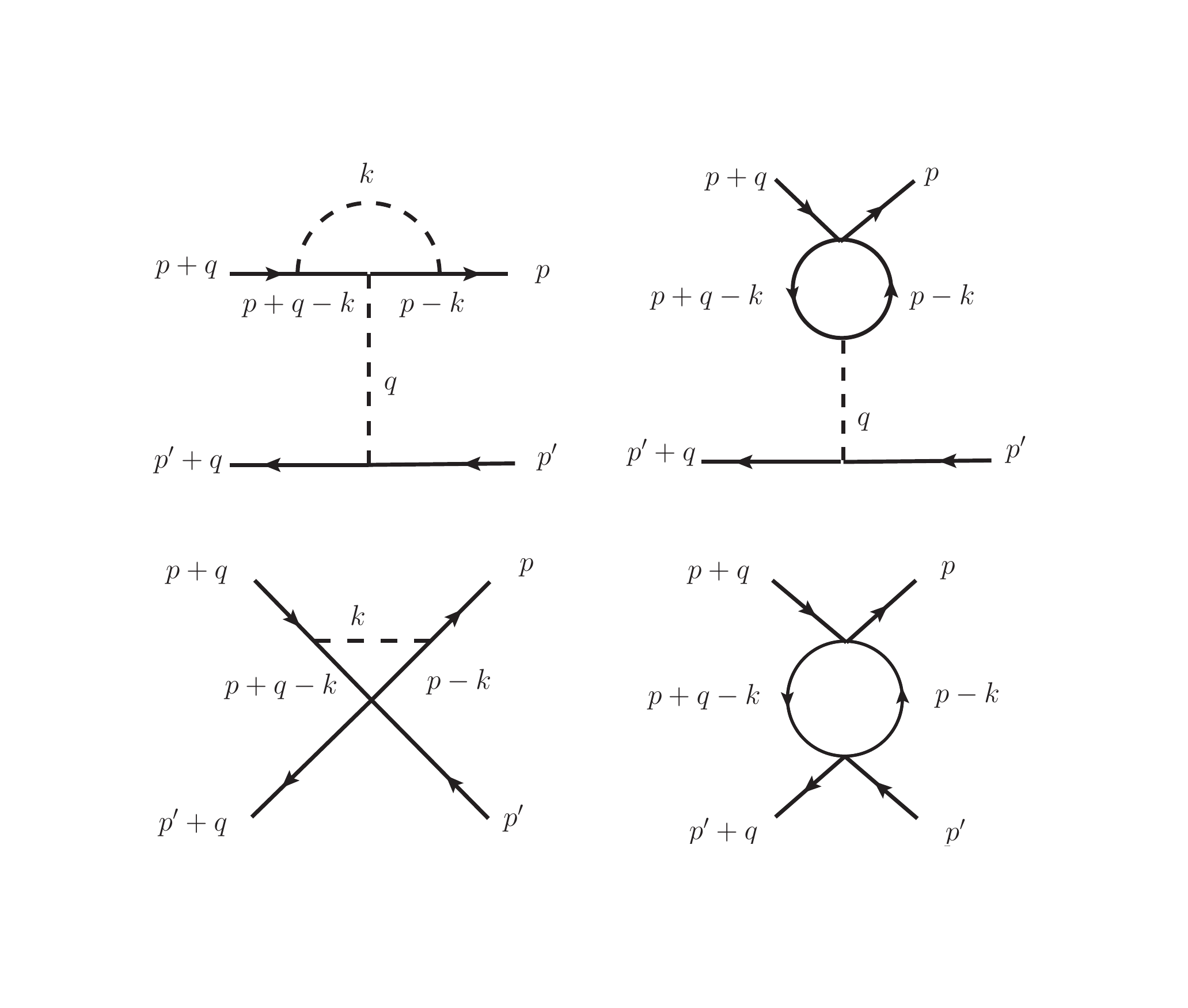}
\end{center}
\caption{\small{Diagrams contributing to the renormalization of the forward scattering Landau parameter at one loop}}\label{fig:Landauparam}
\end{figure}

As the first application of the Wilsonian RG that we have proposed, we will analyze in more detail the renormalization of the 4-Fermi FS coupling. The diagrams that contribute at one loop are shown in Figure \ref{fig:Landauparam}. Note that the zero sound diagrams of Figure \ref{fig:zero-sound} vanish for the forward scattering coupling on the Fermi surface, as the poles from the fermion propagators are always on the same side of the complex plane.

The first diagram in the figure has a one loop vertex correction inserted as a subdivergence in the tree level running of $\lambda$. This will give a logarithmic divergence with both regular and singular terms, as in (\ref{eq:Gr1}) and (\ref{eq:Gs1}). In the theory with a singlet scalar $\phi$ that we have discussed so far, the regular term cancels exactly against the anomalous dimension contribution. However, the anomalous dimension dominates for a theory with matrix-valued $\phi$, where $\psi$ and $\phi$ transform in the fundamental and adjoint representations, respectively. Next, the singular divergence will cancel against the second diagram on top, as in (\ref{eq:Gammatotal}). Finally, combining with the two diagrams in the bottom of the Figure will cancel the $\log \Lambda_b$ divergence from the integral over the boson propagator, replacing it by the physical coupling $\lambda_L$. In summary, in the singlet-scalar $\phi$ theory
\be\label{eq:betaLx}
\beta_{\lambda_L}(x)= g^2+  \frac{g^2}{2\pi^2|v|}\,\frac{1}{1-ix/v}  \lambda_L
\ee
with $x=q_0/q_\perp$. The first term is the tree level running from boson exchange, and the net contribution from the anomalous dimension appears in the matrix-valued $\phi$ case. The origin of the last term is the same as in (\ref{eq:betag}).

This result of our RG approach is important both conceptually and for its possible consequences.
If the interaction is defined right at the Fermi surface, $x \to \infty$ and the one loop correction vanishes. However, in various processes it is natural to consider a static interaction at finite momentum, in which case $x=0$ and we find a net one loop renormalization of the Landau parameter. This same contribution will be obtained in the adiabatic limit $v/c \gg 1$ at fixed $x$. Similar effects can be seen in the BCS channel.

The beta function at $x=0$ also appears to have interesting applications. Indeed, for an attractive interaction we could balance the tree level relevant contribution against the one loop irrelevant term, finding an approximate fixed point at $|\lambda_L|=2\pi^2 |v|$. Of course, this simple analysis will be subject to various corrections (such as those from Landau damping and superconductivity), and it is not our goal here to model-build a controlled fixed point. But the possibility of approximate fixed points for the Landau parameters would lead to interesting novel phases, and we hope to return to this point in future work.

\subsection{Friedel oscillations and the $2 k_F$ vertex}\label{subsec:2KF}

A weakly coupled Fermi liquid has singularities in the density-density correlator $\langle \rho(k) \rho(-k) \rangle$ at $k = 2k_F$. The reason is that a scattering process with large momentum transfer $
\vec K = 2 k_F \hat n$ can take a quasiparticle at the Fermi surface with $\vec p= - k_F \hat n$ into another one with $\vec p + \vec K= k_F \hat n$, which is also at the Fermi surface. These $2k_F$ singularities produce $\sin(2 k_Fr)$ type oscillations in position space, known as Friedel oscillations. The density-density correlator is then an important gauge-invariant probe of the structure of the Fermi surface. This observable also appears in systems with electric impurities interacting with conduction electrons, where it controls the linear response to the presence of impurities.

An important goal is to understand the structure of $2 k_F$ singularities in strongly interacting systems. Previous calculations in $2+1$ dimensions include~\cite{Altshuler1994, Nayak1994a, Mross2010}. A puzzling aspect of the renormalization of the $2 k_F$ response is that double logarithms appear already at one loop. Here we will study the renormalization of the $2k_F$ vertex using our RG approach, and we will see how this problem is resolved. Very similar computations appear in~\cite{ustoappear} for the BCS and CDW instabilities.

\begin{figure}[h!]
\begin{center}
\includegraphics[width=0.8\textwidth]{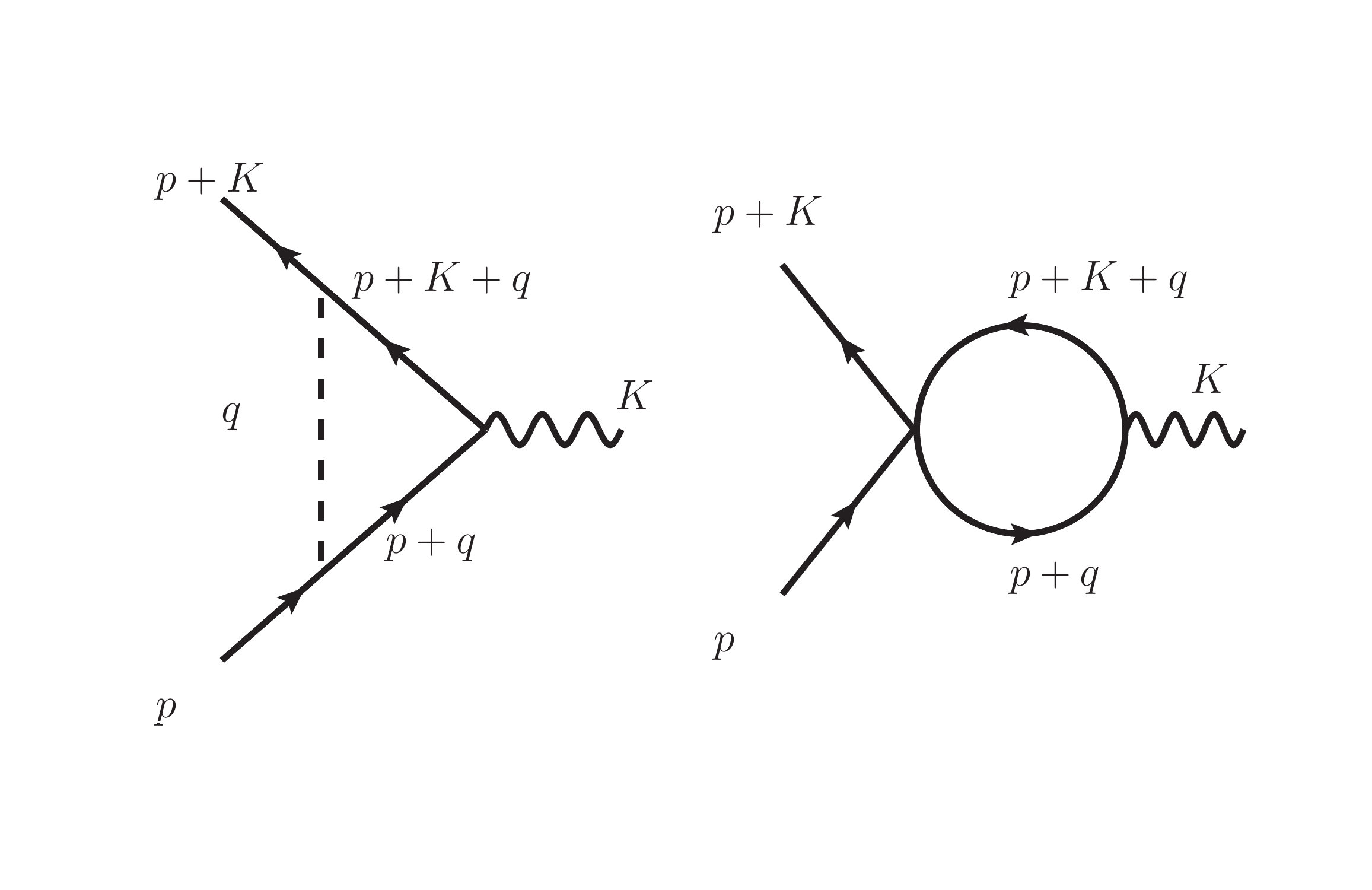}
\end{center}
\caption{\small{One loop renormalization of the $2 k_F$ vertex, represented by the insertion of the wavy line with momentum $|K|=2 k_F$.}}\label{fig:2KF}
\end{figure}

Let us consider the $2 k_F$ vertex in our perturbative theory, which arises from coupling the fermion density to an external gauge field of momentum $K=2 k_F$,\footnote{The overall factor of $-i$ is the Euclidean convention for the $A_0$ component.}
\be
L_{2k_F}= -i  \int\, \frac{d^dp}{(2\pi)^d}\, u_K\, \psi^\dag(p+K) \psi(p)\,.
\ee
 As before, the renormalization is carried out by distinguishing between the bare and renormalized coupling,
\be\label{eq:urenorm}
u_{K,0}= \mu^{1+\epsilon/2}\,\frac{Z_u}{Z_\psi}\,u_K
\ee
where the power of the RG scale is the classical dimension in $d=3-\epsilon$ and $Z_\psi$ accounts for the wavefunction renormalization of the fermions. It is convenient to define $Z_u u_K = u_K+\delta u_K$, and the vertex including quantum effects will be denoted by
\be
\mu^{1+\epsilon/2}(u_K+\delta u_K+\Gamma_K) =-i \langle \psi^\dag(\vec K/2)\psi(-\vec K/2) \rangle\,.
\ee
The one loop contributions to $\Gamma_K$ are shown in Figure \ref{fig:2KF}.

We analyze first the diagram with the virtual boson and $\vec p=-\vec K/2$, given by
\be
\Gamma^{(1)}_K= g^2 u\, \int\,\frac{d^{d+1}q}{(2\pi)^{d+1}}\,D(q)\,\frac{1}{i q_0 - \varepsilon(-K/2+q)}\,\frac{1}{i q_0 - \varepsilon(K/2+q)}\,.
\ee
This has the same structure as the vertex correction of \S \ref{subsec:vertex}, except that the large boson momentum transfer reverses the direction of the Fermi velocity. As a result, the quantum correction will be quite different. Given $\vec K=2 k_F \hat n$, we decompose $ \vec q= q_\perp \hat n + q_\parallel$. Assuming for simplicity a quadratic dispersion relation obtains
\be
\Gamma^{(1)}_K= \frac{g^2 u}{(2\pi)^3}\,\int \frac{dq_0 dq_\perp\, q_\parallel dq_\parallel}{q_0^2+q_\perp^2 +q_\parallel^2}\,\frac{1}{i q_0 +vq_\perp -\frac{v}{2k_F} \vec q^2}\,\frac{1}{i q_0 -vq_\perp-\frac{v}{2k_F} \vec q^2}\,.
\label{eq:friedelint}
\ee

We will analyze the regularization and renormalization of $u_K$, 
and also take the opportunity to spell out in more detail several general aspects of our renormalization prescription. For one, like most of the diagrams in this paper, once the $\frac{v}{2k_F}\vec{q}\ ^2$ terms are included in the fermion propagator, $\Gamma^{(1)}_K$ is UV-convergent as can be seen by power-counting. However, in this case UV-convergence is a drawback since there are large logarithms in the IR that one would like to resum using the RG.\footnote{We thank Catherine Pepin for emphasizing to us this obstacle in some previous treatments.}  This is accomplished by working in the low-momentum effective theory where the $\frac{v}{2k_F}\vec{q}\ ^2$ terms are Taylor expanded and treated in the quadratic fermion action as interaction terms rather than terms that are included in the propagator:
\be
\frac{1}{i q_0 + v q_\perp - \frac{v}{2k_F} \vec{q}\ ^2} \rightarrow \frac{1}{i q_0 + v q_\perp} + \frac{1}{i q_0 + v q_\perp} \frac{v}{2k_F} \vec{q}\ ^2 \frac{1}{i q_0 + v q_\perp}+ \dots
\ee
For the cut-off regulator which is more physically transparent, this just requires taking $\Lambda_b \ll \sqrt{2k_F \mu_f}$ with $\mu_f$ the RG scale  $\mu_f^2 \sim q_0^2+v^2 q_\perp^2$.  At large $k_F$ and fixed $\Lambda_b$, this condition is clearly satisfied.  For the dimensional regulator, we formally take the limit $k_F \rightarrow \infty$ inside the fermion propagator first, and then $\epsilon\rightarrow 0$.  

The integral (\ref{eq:friedelint}) has an infrared divergence that is regulated by taking finite external momenta.  In practice, since we are not specifically interested here in the dependence on this momenta and just want to see the renormalization procedure working, it will be simpler to regulate the infrared by adding small mass terms.  Furthermore, although it is conceptually useful to use a separate cut-off $\Lambda_b$ for bosons and $\Lambda_f$ for fermions, for the example in this section we find it simpler to use a single cut-off $\Lambda > \sqrt{q_0^2 + q_\perp^2 + q_\parallel^2}$ for both.\footnote{It is easy to see that the tree-level log divergence of $\lambda$ is the same with this cut-off as with $\Lambda_b$.}   We will therefore evaluate the log-enhanced parts of the following integral:\footnote{
One way to evaluate this integral is by converting $dq_0 dq_\perp$ to radial coordinates:
\begin{eqnarray}
\Gamma^{(1)}_K &=&  -2 \frac{g^2 u}{(2\pi)^3} \int_0^1 dx x^{-1/2} y^{-1/2} \int_0^\Lambda \rho d \rho \int_0^{\sqrt{\Lambda^2 - \rho^2}} q dq \frac{1}{(\mu^2 + \rho^2(x + v^2 y))(\rho^2 +q^2 + \mu^2)} \nn\\
   &=& -\frac{g^2 u}{(2\pi)^3} \int_0^1 dx \frac{-\text{Li}_2\left(-\frac{\left(y v^2+x\right) \left(\Lambda ^2+\mu ^2\right)}{\left(1-v^2\right) y \mu
   ^2}\right)+\text{Li}_2\left(1-\frac{1}{\left(1-v^2\right) y}\right)+\log \left(\frac{\mu ^2}{\Lambda ^2+\mu ^2}\right) \log
   \left(\frac{1}{\left(1-v^2\right) y}\right)}{2 \sqrt{x y} \left(v^2 y+x\right)} \nn\\
    &\stackrel{\Lambda \rightarrow \infty}{\rightarrow} &- \int_0^1 dx \left[ \log^2 \Lambda \left( \frac{1}{4\sqrt{x y} \left(v^2 y+x\right)} \right) + \log \Lambda \left( \frac{\log \left(\frac{v^2 y+x}{\mu ^2}\right)}{ \sqrt{x y} \left(v^2 y+x\right)}\right) + {\cal O}(\Lambda^0) \right] ,
   \end{eqnarray}
where $y \equiv 1-x$. Performing the $dx$ integration gives (\ref{eq:LambdaTotResult}). 
   $\Gamma^{(2)}_K$ can be evaluated in a similar way to $\Gamma^{(1)}_K$:
\begin{eqnarray}
\Gamma^{(2)}_K &=& 2  \frac{ u }{(2\pi)^3}(\lambda+g^2 \log\left(\frac{\Lambda}{M}\right))\int_0^1 dx x^{-1/2} y^{-1/2} \int_0^\Lambda \rho d \rho  \frac{1}{(\mu^2 + \rho^2(x + v^2 y))} \nn\\
&=& \frac{u }{(2\pi)^3}(\lambda + g^2 \log\left(\frac{\Lambda}{M}\right)) \int_0^1 dx x^{-1/2} y^{-1/2} \frac{1}{x + v^2 y} \log \frac{ (x + v^2 y) \Lambda^2 + \mu^2}{\mu^2},
\end{eqnarray}
and the final integration produces (\ref{eq:LambdaTotResult2}).
   }
\begin{eqnarray}
\Gamma^{(1)}_K &=& -\frac{g^2 u}{(2\pi)^3} \int_{q_0^2 + q_\perp^2 + q_\parallel^2 < \Lambda^2} \frac{ dq_0 dq_\perp q_\parallel dq_\parallel}{(q_0^2 + q_\perp^2 + q_\parallel^2 + \mu^2)(q_0^2 + v^2 q_\perp^2 + \mu^2)}\nn\\
  &=&-\frac{g^2 u}{(2\pi)^2v}  \left( \frac{1}{2} \log^2 \Lambda  -  \log(\frac{1+v}{2v} \mu)  \log \Lambda +{\cal O}(\Lambda^0) \right).
    \label{eq:LambdaTotResult}
 \end{eqnarray}

There is also a contribution from the BCS channel (since $p$ and $p+K$ are nearly opposite) four-Fermi interaction $\lambda_{L=0}^{\text{BCS}}$ and its tree-level counter-term $\delta \lambda_{L=0}^{\text{BCS}} = g^2 \log\left(\frac{\Lambda}{M}\right)$, where $M$ is the RG scale:
\begin{eqnarray}
 \Gamma^{(2)}_K &=&   \frac{ u }{(2\pi)^3}\left(\lambda_{L=0}^{\text{BCS}}+g^2 \log\left(\frac{\Lambda}{M}\right)\right)\int_{q_0^2+q_\perp^2< \Lambda^2}   \frac{dq_0 d q_\perp }{q_0^2 + v^2 q_\perp^2 + \mu^2} \nn\\
 &=& \frac{u }{(2\pi)^2v}\left( g^2\log \frac{\Lambda}{M}\log \Lambda+\lambda_{L=0}^{\text{BCS}} \log \Lambda - g^2\log ( \frac{1+v}{2v} \mu) \log \frac{\Lambda}{M} + {\cal O}(\Lambda^0) \right) .
 \label{eq:LambdaTotResult2}
\end{eqnarray}
We see explicitly that the $\log \mu \log \Lambda$ term cancels in the sum $\Gamma^{(1)}_K + \Gamma^{(2)}_K$ (which is crucial since $\mu$ is playing the role analogous to that  of external momenta):
\begin{eqnarray}
\Gamma^{(1)}_K + \Gamma^{(2)}_K &=& \frac{u}{(2\pi)^2 v} \left( \frac{1}{2} g^2 \log^2 \Lambda+ (\lambda_{L=0}^{\text{BCS}}- g^2 \log(M)) \log \Lambda + {\cal O}(\Lambda^0) \right) \nn\\
 &=& \frac{u}{(2\pi)^2 v} \left( \frac{1}{2} g^2 \log^2 \frac{\Lambda}{M} + \lambda_{L=0}^{\text{BCS}} \log \frac{\Lambda}{M} + {\cal O}(\Lambda^0) \right).
\end{eqnarray}
From this, we read off that the counter-term for $u$ at this order must be
\be
\delta u = -\frac{u}{(2\pi)^2v } \left( \frac{1}{2} g^2 \log^2 \frac{\Lambda}{M} + \lambda_{L=0}^{\text{BCS}} \log \frac{\Lambda}{M}\right) + \textrm{finite},
\ee
where the finite $\Lambda$-independent piece is scheme-dependent and we will choose it to vanish.  

The $\beta$ function for the dimensionless coupling $u$ can be determined by the condition that the bare term $u_0 =M( u+\delta u- u \delta Z_\psi)$ is independent of the RG scale $M$:
\be
0 = M \frac{d}{dM} u_0  = M \frac{d}{dM} (M(u+ \delta u- u \delta Z_\psi)) =u+ \beta_u - 2 \gamma_\psi u+ M \frac{d}{dM} \delta u,
\ee
where $\beta_u = M \frac{d}{dM} u$.   Collecting terms in an expansion in powers of $\log\frac{\Lambda}{M}$, we therefore have
\begin{eqnarray}
0 &=& \left(u+\beta_u -2 \gamma_\psi u + \frac{u \lambda_{L=0}^{\text{BCS}}}{(2\pi)^2 v}\right) + \log\frac{\Lambda}{M} \frac{u}{(2\pi)^2 v}\left( -g^2 +\beta_\lambda+ \dots \right) + \dots
\label{eq:LogExpansion}
\end{eqnarray}
where `$\dots$' denotes higher orders in $\log\frac{\Lambda}{M}$ and/or couplings.  We can read off the $\beta_u$ function from the cancellation of the $\Lambda$-independent piece, 
\be\label{eq:beta2kFfinal}
\beta_u=  u \left(-1+ 2 \gamma_\psi - \frac{\lambda_{L=0}^{\text{BCS}}}{(2\pi)^2 v} \right),
\ee
and the cancellation in the coefficients of the higher powers of $\log\frac{\Lambda}{M}$ provides consistency conditions that will be satisfied due to the running from lower-order diagrams. In particular, we see from the linear in $\log(\frac{\Lambda}{M})$ term in (\ref{eq:LogExpansion}) that $\beta_\lambda=g^2$.  So we see here that even if we had not thought to consider tree-level running of $\lambda$, we would have noticed it must be included just from analysis of the diagrams in Figure \ref{fig:2KF}.

Eq.~(\ref{eq:beta2kFfinal}) is our final result for the renormalization of the $2k_F$ vertex, showing how multilogarithms are properly taken into account by the running of the couplings.
The fermion anomalous dimension  (\ref{eq:beta1}) always makes the vertex irrelevant, and the same is true for an attractive $\lambda_{L=0}<0$ interaction (generated e.g. by the exchange of high momentum bosons).
We conclude that in systems with gapless spin zero bosons, perturbative quantum corrections tend to smooth out the $2 k_F$ singularities. In contrast, the sign of the 4-Fermi contribution is reversed in a theory with gauge fields,\footnote{The gauge field, being a vector, couples with opposite signs to fermions with opposite Fermi velocity. This is unlike the scalar field, which couples with the same sign to all patches. This was also observed in~\cite{Mross2010}.} so there can be a competition between the vertex correction and $\gamma_\psi$. On a different note, holographic models at finite density have signatures of Fermi surfaces that are strongly suppressed~\cite{Polchinski:2012nh, Faulkner:2012gt}, and it is interesting that our perturbative results also point in the same direction. It would be important to try to continue this calculation to strong coupling, and to apply these results to models with impurities.

\section{Renormalization at higher orders and Landau damping}\label{sec:higher}

So far we have established the renormalizability of the theory of a Fermi surface coupled to a massless boson at one loop order, and have set up a consistent (in the sense of including the dominant quantum effects) Wilsonian RG framework. We expect that our approach for organizing divergences and the RG can also be extended to higher loop level without obstruction, though at present we have no general proof. A new element to take into account is that when going to higher orders it is necessary to include Landau damping. We will now explain how to extend our RG to include such effects.\footnote{Since a detailed analysis of Landau damping and how it affects fermion correlation functions was recently performed in~\cite{Torroba:2014gqa}, our discussion here will be brief.}

\begin{figure}[h!]
\begin{center}
\includegraphics[width=0.9\textwidth]{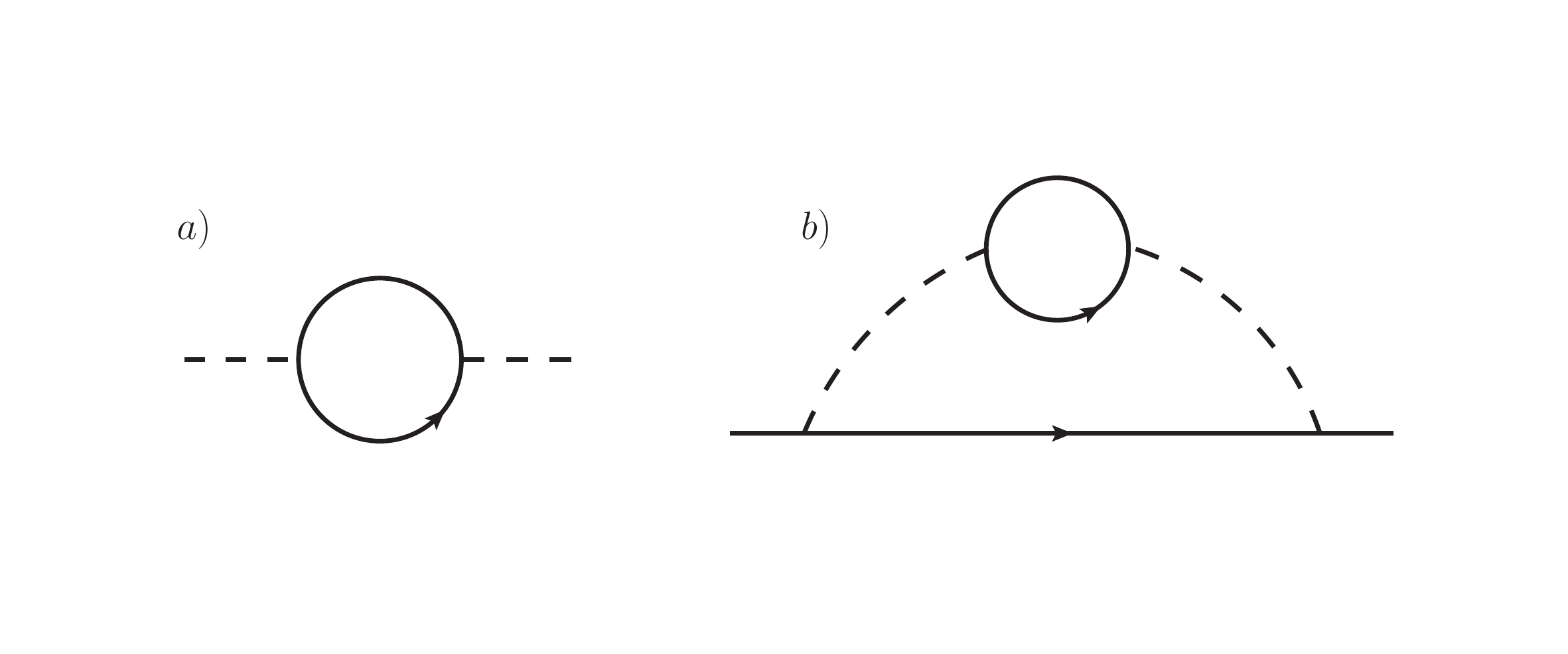}
\end{center}
\caption{\small{a) Landau damping for the boson self-energy. b) Two loop correction to the fermion self-energy with a boson self-energy insertion.}}\label{fig:damping}
\end{figure}

Landau damping effects come from the one loop fermion bubble that contributes to the boson self-energy, Fig.~\ref{fig:damping} (a). This diagram is finite, and that is why formally it did not affect the one loop RG; crucially, however, the large Fermi momentum $k_F$ reappears here when integrating over the Fermi surface, with the result that (a) becomes important at a high scale proportional to $g k_F$. This in turn gives two loop corrections to the RG, such as diagram (b) in Fig.~\ref{fig:damping}, that are comparable or can even dominate over the one loop results. At this point, the standard perturbative RG breaks down. The failure of the perturbative expansion is due to the fact that the loop factor $g^2/16\pi^2$ from the additional boson self-energy insertion in (b) also comes with a power of $k_F^2$ from the integral over the Fermi surface.

A well-known solution to this problem is to reorganize the perturbative expansion in terms of the resummed boson propagator,
\be\label{eq:Ddamped}
D^{-1}(p) = p_0^2+ \vec p^{\,2} + \Pi(p)\;,\;\Pi(p)= M_D^2\,\frac{p_0}{v|\vec p\,|}\,\tan^{-1} \frac{v |\vec p\,|}{p_0}\,,
\ee
with
\be
M_D^2= \frac{g^2 k_F^2}{2\pi^2 v}\,,
\ee
and $\Pi(p)$ is the result of evaluating diagram (a). Physically, Landau damping comes from virtual particle/hole pairs near the Fermi surface, whose contribution is given by integrating the delta function peak (\ref{eq:GG}) over the Fermi surface (up to a constant term that needs to be fixed in order to tune the boson to criticality). We have seen before that for consistency the RG needs to include this region near the Fermi surface (see e.g.~Fig.~\ref{fig:RGstep}), and it is satisfying that this same prescription also captures Landau damping. Furthermore, note that
at low energies the propagator can be approximated by
\be\label{eq:z3boson}
D^{-1}(p)\approx \vec p^{\,2}+\frac{\pi}{4}M_D^2\,\frac{p_0}{v|\vec p\,|}\,,
\ee
giving a boson with $z=3$ dynamical exponent \cite{Hertz1976, Millis1993}.

From the point of view of the original theory, this resummation amounts to including an infinite class of diagrams, and it is important to determine when this is consistent. Within our perturbative framework with small coupling $g$ and near three dimensions this appears to be the case. Examination of diagrams reveals that (a) in the figure gives the dominant nonanalytic contribution responsible for the $z=3$ scaling, and that other effects are perturbative analytic corrections on this. It would be interesting to understand this more generally, but here we will assume (\ref{eq:Ddamped}) and study its consequences.

Let us for simplicity restrict to scales much smaller than
\be
\mu_{LD} \approx \left(v^{-1}\,\tan^{-1}(v) \right)^{1/2}\, M_D\,,
\ee
where (\ref{eq:z3boson}) is a good approximation. The one loop fermion self-energy and vertex corrections (diagrams $\Sigma_a$ and $\Gamma_a$ in Fig.~\ref{fig:diagrams1}) using the Landau damped propagator give~\cite{Torroba:2014gqa}
\be\label{eq:Sigma-resummed}
\Sigma_a(q)\approx i\,\frac{g^2}{12\pi^2|v|}\,q_0\log\frac{\Lambda}{q_0}\,,
\ee
as well as a nonlocal divergence
\be
\Gamma_a(k;q) =\frac{g^3}{12\pi^2|v|}\,\frac{iq_0}{iq_0-v q_\perp}\,\log\frac{\Lambda}{q_0}\,.
\ee
The factor of $3$ difference with what we obtained above in \S \ref{sec:quantum} is a consequence of the boson dynamical exponent. Here we have for simplicity considered a single cutoff $\Lambda$, with $\Lambda_b^3 \propto \Lambda_f \sim \Lambda$. The tree level running similarly becomes $\beta_{\lambda_L}=g^2/3$.

The renormalization now proceeds as before, by including the one loop contributions $\Sigma_b$ and $\Gamma_b$ from the 4-Fermi interaction in Fig.~\ref{fig:diagrams1}. In particular, the fermion self-energy becomes
\bea
\Sigma(q)&\approx &\frac{g^2}{12\pi^2|v|}\,\log\Lambda\,(i q_0-v q_\perp)-\frac{\lambda_{L=0}}{4\pi^2|v|}\,v q_\perp \nonumber\\
\Gamma(q)&\approx &\frac{g^3}{12\pi^2|v|}\,\log\Lambda\,-\frac{g\lambda_{L=0}}{4\pi^2|v|}\frac{v q_\perp}{i q_0 -vq_\perp}
\eea
In the overdamped regime the couplings always appear in combinations $g^2/|v|$ and $\lambda_L/|v|$ (as can be seen by redefining the fields), so it will be convenient to define
\be
\alpha\equiv \frac{g^2}{12\pi^2|v|}\,,\,\tilde \lambda_L \equiv \frac{\lambda_L}{|v|}\,.
\ee
Using the physical renormalization scheme of \S \ref{sec:quantum} obtains the following beta functions:
\bea\label{eq:dampedbeta}
\gamma_\psi&=& \frac{\alpha}{2}\;,\;\beta_v=\alpha v\nonumber\\
\beta_{\tilde \lambda_L}&=&4\pi^2 \alpha- \alpha\,\tilde \lambda_L +\frac{2\alpha}{1-ix/v}\,\tilde \lambda_L\\
\beta_\alpha&=&-\epsilon \alpha -\alpha^2+\frac{2\alpha^2}{1-ix/v}\nonumber\,.
\eea
The origin of the $x=q_0/q_\perp$ dependent terms is the same as in (\ref{eq:betag}) and (\ref{eq:betaLx}).

For the renormalization condition $x \to \infty$ these beta functions do not admit fixed points. However, if $x=0$ the one loop $\beta$ functions for $\alpha$ and $\tilde{\lambda}_L$ have zeros at $\alpha=\epsilon$, $\tilde \lambda_L=-4\pi^2$. The later is not under perturbative control, and it would be interesting if one could find an exact solution, or
models where a small $\tilde \lambda_L$ can be achieved. 

Finally, we note that in a large $N$ generalization where $\phi$ is an $N \times N$ matrix and $\psi$ is a vector, the $x$ dependence is suppressed, and the $ \alpha\,\tilde \lambda_L$ and $\alpha^2$ terms change sign. This parameter range gives the possibility of flowing to a non-Fermi liquid fixed point \textit{before} reaching the superconducting scale, something that was briefly studied in~\cite{Torroba:2014gqa} and that we plan to analyze in more detail in the future.

\section{Future directions}\label{sec:future}

In this work we have studied finite density QFT and established its renormalizability at one loop. We focused on the important example of a Fermi surface coupled to a gapless boson mode, and established an RG procedure that consistently takes into account the dominant quantum corrections. The key features of this approach are the tree level running of all 4-Fermi interactions, together with a cutoff prescription that includes the UV-IR mixing of bosons and fermions. We also discuss how to partially extend the approach to higher loop order, by adding Landau damping to the RG.

Our results provide a framework where quantum corrections at finite density can be systematically calculated and incorporated to the RG. It would clearly be important to prove the renormalizability at all orders, perhaps generalizing~\cite{Polchinski:1983gv} to finite density. It would also be interesting to apply our methods to the theory introduced in~\cite{Mandal:2014cma}, with dimensional regulators $\epsilon_\perp$ and $\epsilon_\parallel$ in both perpendicular and parallel directions to the Fermi surface. We anticipate nontrivial renormalization effects at finite $\epsilon_\perp$. Finally, we have seen the possibility of fixed points for 4-Fermi interactions. Such fixed points could lead to novel IR phases, and in future work we plan to explore this direction in more detail.

\section*{Acknowledgments}
We thank S. Kachru, J. Kaplan, M. Mulligan and S. Raghu for interesting discussions about related
subjects. GT is supported by CONICET, and PIP grant 11220110100752. H.W.~ is supported by a Stanford
Graduate Fellowship.

\appendix

\section{Dimensional regularization at finite density}\label{app:dimreg}

In this Appendix we will study finite density QFT in dimensional regularization, with $d=3-\epsilon$. This has the effect of analytically continuing the number of tangential directions to the Fermi surface, $d_\parallel=2-\epsilon$. We note that this is different from the dimensional regulator of~\cite{Senthil2009, Lee2013}, where the normal directions to the Fermi surface are fractional.

\subsection{Vertex and fermion self-energy corrections}

Let us begin with 
the one loop vertex correction $\Gamma_a$ of Figure \ref{fig:diagrams1} in DR,
\be\label{eq:Vertex_correction2}
\Gamma_a(k;q 
)\approx\mu^\epsilon\frac{g^3}{(2\pi)^{d+1}}\int\frac{dp_0\,dp_\perp\,d^{d-1}
p_\parallel}{(k_0-p_0)^2+(k_\perp - p_\perp)^2+p_\parallel^2}\,\frac{1}{ip_0-v 
p_\perp}\,\frac{1}{i(p_0+q_0)-v(p_\perp+q_\perp)}\,.
\ee

The regular term becomes
\be
\Gr=-\frac{g^3}{8\pi^2 \epsilon}\,\frac{1-|v|}{|v|(1+|v|)}\,,
\ee
and we can also calculate directly the total contribution~\cite{Torroba:2014gqa}
\be\label{eq:dimreg_gamma}
\Gamma_a(\hat n 
k_F;q)=\frac{g^3}{4\pi^2}\frac{1}{1+|v|}\frac{iq_0+\text{sgn}(v) 
q_\perp}{iq_0-v q_\perp}\frac{1}{\epsilon} + \mc O(\epsilon^0)\,.
\ee
From here, the singular term is given by the difference
\be\label{eq:Gsdimreg}
\Gs=\Gamma-\Gr=\frac{g^3}{8\pi^2|v| \epsilon}\,\frac{iq_0+v q_\perp}{iq_0-v q_\perp}\,.
\ee

Notice that we can use this result for $\Gs$ to define how to evaluate $\int_p G(p) G(p+q)$ in DR, which by itself is not regulated by $\epsilon$. We will need this expression in order to calculate $\Gamma_b$. As explained in the main body of the text, for small $q$ it is enough to consider an ansatz
\be\label{eq:GGansatz}
G(p) G(p+q) \approx G(p)^2 + f(q)\, \delta(p_0) \delta (p_\perp)\,.
\ee
By definition, $\Gs$ is the piece that comes from this delta function term. Integrating (\ref{eq:GGansatz}) and requiring that $f(q)$ reproduces $\Gs$ gives
\be\label{eq:GGDR}
G(p) G(p+q) \approx G(p)^2 +\frac{\pi i}{|v|}\,\frac{i q_0+v q_\perp}{i q_0-v q_\perp}\delta(p_0) \delta (p_\perp)\,.
\ee
This should be contrasted with the cutoff prescription (\ref{eq:GG}). From this point of view, DR corresponds to averaging the results of integrating over $p_0$ first, and integrating over $p_\perp$ first, something that makes sense as these coordinates are not distinguished.

With (\ref{eq:GGDR}) we can now evaluate $\Gamma_b$:
\be
\Gamma_b=-\frac{g}{8\pi^2 |v|}(\lambda_{L=0}+\delta\lambda_{L=0})\,\frac{i q_0+v q_\perp}{i q_0-v q_\perp}\,.
\ee
Therefore, in DR the cancellation of divergences between $\Gamma_a$ and $\Gamma_b$ works out to give
\be\label{eq:Gammatotal2}
\Gamma(p;q)\approx -\frac{g^3}{8\pi^2 \epsilon}\,\frac{1-|v|}{|v|(1+|v|)}-\frac{g\,\lambda_{L=0}}{8\pi^2|v|}\,\frac{iq_0+v q_\perp}{iq_0-v q_\perp}\,.
\ee
The one loop divergent contributions to $\Gamma$ are hence different in the cutoff approach and in dimensional regularization, something to be expected on general grounds. Indeed, due to the tree level running, scheme dependence in finite density QFT with multiple couplings will occur at one loop. In our case, the divergent piece in dimensional regularization is somewhat unphysical, due to the apparent divergence for $v \to 0$ (which is cancelled by the finite piece). The full correlator does not diverge in this limit, something that is explicit also in the cutoff result (\ref{eq:Gammatotal}). In a physical subtraction scheme the cutoff and dimensional regulators will give the same results, as we verify shortly.

Let us now discuss the corrections to the fermion self-energy.
The first contribution $\Sigma_a$ in Figure \ref{fig:diagrams1} 
gives~\cite{Fitzpatrick:2013rfa,Torroba:2014gqa}
\bea\label{eq:self-energy2}
\Sigma_a(k_0, \vec k) &\approx&-g^2 \mu^\epsilon\,\int \frac{dp_0\, d p_\perp 
\,d^{d-1} 
p_\parallel}{(2\pi)^{d+1}}\,\frac{1}{p_0^2+p_\perp^2+p_\parallel^2}\,\frac{1}{
i(p_0+k_0)- v (p_\perp + k_\perp)}\nonumber\\
&=&\frac{g^2}{4\pi^2(1+|v|)}\left(i k_0 + 
\text{sgn}(v)k_\perp\right)\frac{1}{\epsilon} + \mc O(\epsilon^0)\,,
\eea
with momenta $\vec k = \hat n(k_F + k_\perp)$, $\vec p = \hat n p_\perp + \vec 
\pp$.  We see that (\ref{eq:dimreg_gamma}) and (\ref{eq:self-energy2}) satisfy 
the Ward identity (\ref{eq:Ward2}). From this point of view, the nonlocal term 
in $\Gamma_a$ is equivalent to a renormalization of the velocity, namely a 
self-energy that does not depend only on $i k_0 - v k_\perp$.

For $\Sigma_b$ we will introduce separate dimensional regularization parameters 
$\epsilon_\parallel$ and $\epsilon_\perp$ for the dimension and codimension, 
respectively, of the Fermi surface.  The dimension of space is then 
$d=3-\epsilon_\perp-\epsilon_\parallel$.  When we take the $\epsilon 
=\epsilon_\perp+\epsilon_\parallel \rightarrow 0$ limit, we shall first take 
$\epsilon_\perp\rightarrow 0$ followed by $\epsilon_\parallel \rightarrow 0$.  
As discussed in the main text, this leaves a finite shift ambiguity of the form (\ref{eq:ambiguity1}). The Ward identity then fixes $a_\perp$, yielding
\be\label{eq:Sigmab2DR}
\Sigma_b(k)=-\frac{\lambda_{L=0}+\delta \lambda_{L=0}}{8\pi^2|v|}\,(i k_0+v k_\perp)\,.
\ee
The final DR result for the self-energy is
\be\label{eq:SigmatotalDR}
\Sigma(k)\approx -\frac{1}{8\pi^2|v|}\left(\frac{1-|v|}{1+|v|}\,\frac{g^2}{\epsilon}+\lambda_{L=0} \right)\,(ik_0- 
vk_\perp)-\frac{\lambda_{L=0}}{4\pi^2}\,\sv\,k_\perp\,.
\ee
As noted above, in this scheme the apparent divergence at small velocities is cancelled between the $1/\epsilon$ and finite piece.

\subsection{$\beta$ functions}

In DR, the simplest subtraction scheme is minimal subtraction (MS), where the counterterms are defined to cancel only the 
$\epsilon$ pole. However, this scheme leads to unphysical results --in particular, from (\ref{eq:SigmatotalDR}) it leads to an anomalous dimension with the wrong sign. This is an artifact, which originates in neglecting the contribution from the physical coupling. 

This motivates adopting a physical subtraction scheme for DR, as we did with the cutoff regulator. Now the counterterms become
\be
\delta_\psi= \frac{1}{8\pi^2 |v|}\left(\frac{1-|v|}{1+|v|}\,\frac{g^2}{\epsilon}+\lambda_{L=0} \right)\;,\;\delta v= 
-\frac{\sv\,\lambda_{L=0}}{4\pi^2}
\ee
and we reproduce the anomalous dimension and running velocity (\ref{eq:beta1}).
Similarly, from (\ref{eq:Gammatotal2}),
\be
\delta 
g=-\frac{g^3}{8\pi^2 }\,\frac{1-|v|}{|v|(1+|v|)}\,\log\mu-\frac{g\,\lambda_{L=0}}{8\pi^2|v|}\,\frac{1+ix/v}{1-ix/v}\,,
\ee
at a renormalization point $q_0 = x \mu\,,
q_\perp=\mu$. The beta function computed from here agrees with the cutoff result (\ref{eq:betag}).



\bibliography{FDRG}{}
\bibliographystyle{utphys}
\end{document}